\documentclass[11pt]{article}
\usepackage{enumerate}
\begin{document}
\input epsf
\def\be{\begin{equation}}
\def\bea{\begin{eqnarray}}
\def\ee{\end{equation}}
\def\eea{\end{eqnarray}}
\def\d{\partial}
\def\eps{\varepsilon}
\def\la{\lambda}
\def\b{\bigskip}
\def\r{\rightarrow}
\def\nn{\nonumber \\}
\def\p{\partial}
\def\h{{1\over 2}}
\def\t{\tilde}
\def\da{\downarrow}
\def\ua{\uparrow}

\vspace{20mm}
\begin{center}
{\LARGE  A toy  black hole S-matrix in the D1--D5 CFT}

~\\
\vspace{20mm}
{\bf  Oleg Lunin${}^1$ and  Samir D. Mathur${}^2$\\}

\vspace{4mm}
${}^1${\it{Department of Physics\\ University at Albany (SUNY)\\ Albany NY 12222}
\bigskip
 
${}^2$Department of Physics,\\ The Ohio State University,\\ Columbus, OH
43210, USA\\
}
\vspace{4mm}
\end{center}
\vspace{10mm}
\begin{abstract}

To model the process of absorption and emission of quanta by an extremal D1--D5 black hole in the dual CFT, we consider transitions between different Ramond vacua via absorption and emission of chiral primaries. We compute the probabilities to reach different CFT states starting with a special Ramond vacuum, using techniques of the orbifold CFT.  It is found that the processes involving the 
change of angular momentum by $k$ units are suppressed as  
$\sim 1/N^k$.

\end{abstract}

\newpage

\section{Introduction}
\label{SectIntro}
\renewcommand{\theequation}{1.\arabic{equation}}
\setcounter{equation}{0}

Black hole formation and evaporation is expected to be a unitary 
process, but the details of this process are not fully understood, in particular, it is not clear how information 
comes out of the hole. Consider a simple process
depicted in Fig.1(a): a massless quantum falls into an extremal black hole 
and excites it, and at a later time another massless quantum is radiated
away, bringing the hole back to extremality. Since the extremal hole 
has a large number of ground states, generically the final state $|f\rangle$
of the hole will differ from the initial state $|i\rangle$, and the emitted quantum will differ from the absorbed one. This leads 
to an S matrix with a large number of nonzero elements, and by understanding properties of this matrix, one would shed light on dynamics of non--extremal black holes and on physics of Hawking radiation 
\cite{hawking}. In particular, it would be very interesting to know whether a given initial state $|i\rangle$ 
tends to go to some specific final state $|f\rangle$ or all outcomes 
happen with comparable probabilities. 

In this paper we are focusing on processes depicted in Fig.1(a) for the case of extremal five--dimensional black holes composed of $n_1$ D1 and $n_5$ D5 branes \cite{stromvafa}.
Such system has $e^{2\sqrt{2} \sqrt{N}}$ degenerate ground states, where $N=n_1n_5$. The geometries corresponding to all such states have been constructed in \cite{lmBH,lmm,SkTay}, and although 
these metrics are regular \cite{lmm}, the curvature can become large, so supergravity approximation cannot be used to give an accurate description of the process depicted in Fig.1(a). Moreover, if the initial and final microstates are not the same, such process cannot be described by propagation of a graviton on a fixed background. 
However, some insights into the absorption/emission process can be gained from going to a different regime of parameters, where the system has a dual description in terms of a two--dimensional CFT \cite{AdSCFT}. In this theory, the relevant  process is given by the 4-point function 
depicted in Fig.1(b), and we will compute the appropriate correlators for special cases when the initial state $|i\rangle$ 
corresponds to the geometry constructed in \cite{mm,LMmetr}, and the final state $|f\rangle$ is a global rotation of $|i\rangle$ (however, in general, $|f\rangle\ne |i\rangle$). For this class of transitions, we find that the amplitude for going to a state rotated by $k$ units (i.e., 
$|f\rangle=(J_0^-)^k|i\rangle$) is suppressed by a
factor $1/N^k$ (note that $1/N$ is the effective gravitational constant for the geometry). 

For the D1--D5 system exact agreement between gravity and the free CFT has been found 
for quantities pertaining to the extremal and to the near 
extremal systems\footnote{One can regard the 3-charge hole as an excited state
of the 2-charge D1-D5 system, and thus work around extremal D1-D5 
states.} \cite{callmald,dasmathur,MaldStrom,Klebanov}. Moreover, this agreement persisted beyond supergravity: correlation functions in string theory on AdS$_3\times $S$^3$ \cite{GKDP} turned out to be equal to the ones computed in the free CFT \cite{lm,lmtwo}. It is possible that the low energy S-matrix describing
processes shown in Fig. 1(a) would also agree between the free CFT and the gravity regimes. In this case many interesting properties of the black hole S--matrix can be deduced by studying simple processes similar to the one discussed in this paper.

This paper has the following organization. In section \ref{SectTwist} we review the construction of operators in the orbifold CFT. Section \ref{SectSmatr} presents an outline of the CFT calculation, which is carried out in sections \ref{SectGen4ptRed}--\ref{SectBasic}. The results are summarized in section \ref{SectResult}, and section \ref{SectGrav} describes the implications for the gravitational amplitudes. The technical details are presented in the appendices.

%==============================================================
%                   Cylinder BH figure
%==============================================================
\begin{figure}
\begin{tabular}{cc}
\begin{tabular}{l}
\begin{picture}(85.00,15.00)
\end{picture}\\
\epsfysize=0.8in \epsffile{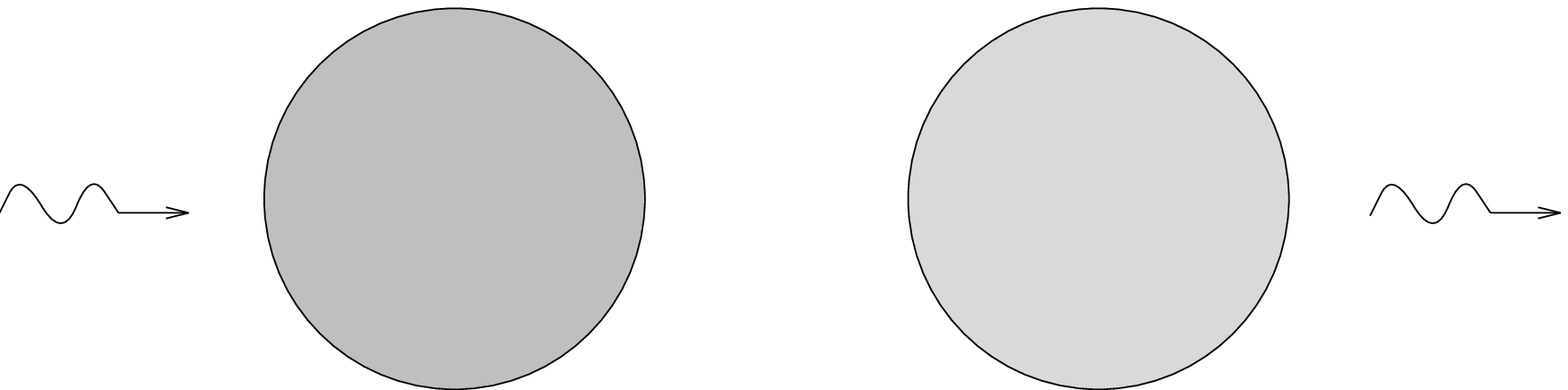}\\
\begin{picture}(225.00,15.00)
\put(162,8){$|f\rangle$}
\put(62,8){$|i\rangle$}
\put(7,60){$|i'\rangle$}
\put(213,60){$|f'\rangle$}
\end{picture}\\
\end{tabular}
&
\begin{tabular}{rl}
\begin{picture}(85.00,15.00)
\put(47,10){$|f\rangle$}
\end{picture}&\\
\epsfysize=1.5in \epsffile{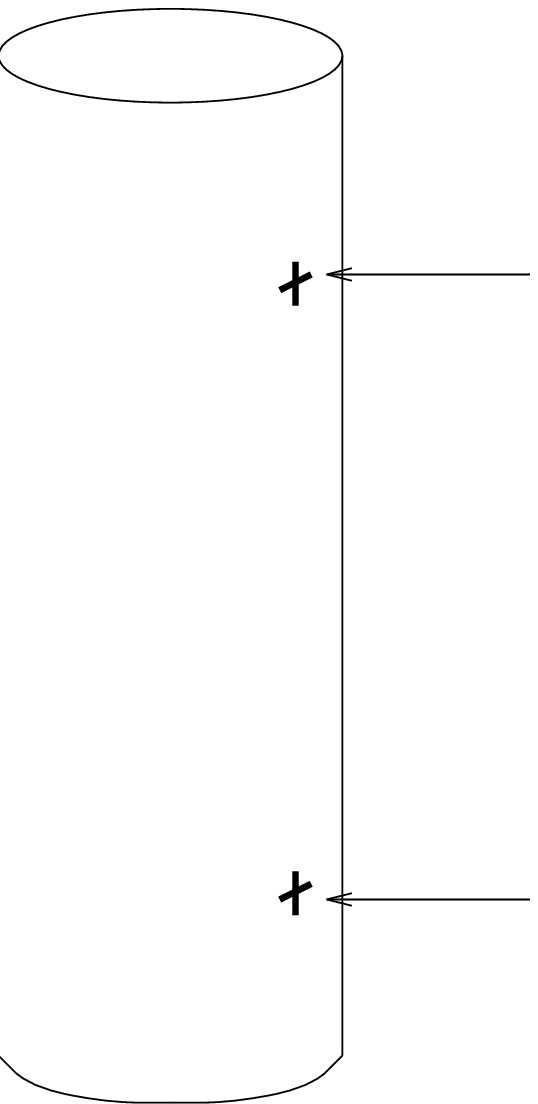}&
\begin{picture}(70.00,80.00)
\put(-5,17){$|i'\rangle$}
\put(-5,80){$|f'\rangle$}
\end{picture}\\
\begin{picture}(85.00,15.00)
\put(47,0){$|i\rangle$}
\end{picture}&
\end{tabular}\\
(a)&(b)\\
\end{tabular}
\caption{\label{figGround}\newline (a) Absorption of a quantum by a black hole (left) and emission of another quantum at a later time (right).\newline
(b) Representation of the same process in the dual CFT. Time is flowing upwards.}
\end{figure}
%==============================================================

\section{Operators in the $M^N/S_N$ CFT }
\label{SectTwist}
\renewcommand{\theequation}{2.\arabic{equation}}
\setcounter{equation}{0}

To evaluate the correlation function mentioned in the introduction, one should begin with identifying the states in the field theory 
which correspond to the black hole and to the absorbed and emitted particles. In this section we will review the construction of such 
states. 

The D1-D5 system is believed to have an `orbifold point' in its 
moduli space where the low energy theory  is a 1+1 dimensional sigma
model with target space $M^N/S_N$ \cite{stromvafa,sw} -- the symmetric product of  $N$ 
copies of a 4-manifold $M$ (which can be either $T^4$ or $K3$).  
A general method for computing correlation functions in this CFT was 
developed in \cite{lm, lmtwo}, and we will begin with reviewing this construction. 

Each of the $N$ copies of the CFT has 4 free bosons $\phi_1, \dots \phi_4$ and
4 free fermions $\psi_1, \dots \psi_4$, so its central charge is $c=6$. 
The fermions can be bosonized into two additional bosons,
which will be denoted by
$\phi_5$ and $\phi_6$. The theory has $N=(4,4)$ supersymmtery, and the R--symmetry in the holomorphic sector is 
generated by $SU(2)$ currents:
\be\label{generatorS}
J^3(z)=\frac{i}{2}\sum_{j}e_a\partial_z \phi^a_j(z),~~~
J^+(z)=\sum_{j}\exp\left(ie_a\phi^a_j(z)\right),~~~
J^-(z)=\sum_{j}\exp\left(-ie_a\phi^a_j(z)\right)
\ee
The six--dimensional vector $e_a=(0,0,0,1,-1)$ was defined in \cite{lmtwo}, and we will only need the relation
$e_a e_a =2$.

\subsection{Chiral primaries}

The incoming and outgoing supergravity quanta in Fig 1 correspond to chiral primaries and their descendants in the 
dual CFT.  The chiral primaries in the orbifold theory have been discussed in \cite{lmtwo}, and here we summarize the results. 

To construct a chiral primary, one starts with a twist operator $\sigma_n$, which interchanges $n$ different copies of the free CFT with $c=6$. Such twist can be resolved by going to a `covering space' $\Sigma$, via a 
map that behaves like $z\approx b t^n$ at the insertion\footnote{In this paper we follow notation of \cite{lm,lmtwo} and denote 
the coordinate on the 1+1 dimensional base space by $z$, and the coordinate on the cover $\Sigma$ by $t$.}. To make a chiral primary with $j=h$, one should add an SU(2) charge to the twist operator by applying  currents $J^+$.  Because of the twist, one can apply 
fractional modes of these currents in the $z$ plane,
\be
J^{+(z)}_{-m/n}\equiv \int {dz\over 2\pi i} \sum _{k=1}^n
~J_z^{k,+}(z) ~e^{-2\pi i
m(k-1)/n} ~z^{-m/n},
\label{six}
\ee
these modes become integral in the covering space:
\bea
J^{+(z)}_{-m/n}=\int {dz\over 2\pi i} \sum _{k=1}^n ~J_z^{k,+}(z)
~e^{-2\pi i m(k-1)/n}
~z^{-m/n}\rightarrow \int{ dt\over 2\pi i}  ~J_t^{+}(t)
~a^{-m/n} t^{-m}\equiv a^{-m/n}~J_{-m}^+.\nonumber
\eea
A chiral primary $\sigma_n^p$ is obtained by applying a sequence of $J^{+(z)}_{-m/n}$ to the bosonic twist operator 
$\sigma_n$, and details of this construction are given in \cite{lmtwo}. The resulting operator has charges 
$j=h\equiv p=\frac{n\pm 1}{2}$, and on the covering space is it described by an insertion of an 
exponential\footnote{In \cite{lmtwo} the two possibilities 
for $p$ were
denoted $+$ and $-$. There is a similar charge for the 
antiholomorphic sector, giving four possibilities overall; we will 
however talk
about the holomorphic sector alone for most of the computation, and 
combine sectors at the end.}:
\be\label{DefCoverTwist}
\sigma^{P}_n(0)\rightarrow
{\hat\sigma}^{P}_n(t=0)=b^{-p^2/n}:\exp\left(ipe_a\Phi^a(0)\right).
\ee

%Applying such modes of $J^+$ we obtain from a twist $\sigma_n$ a 
%chiral primary $\sigma_n^p$, where $p$ if the $J^3$ charge and can be
%${n-1\over 2}$ or ${n+1\over 2}$. Details of this construction are 
%given in \cite{lm2}.\footnote{In \cite{lm2} these two possibilities 
%for $p$ were
%denoted $-, +$ respectively. There is a similar charge for the 
%antiholomorphic sector, giving four possibilities overall; we will 
%however talk
%about the holomorphic sector alone for most of the computation, and 
%combine sectors at the end.}
%If the chiral primary $\sigma^{p}_n $ inserted at $z=0$ and the map 
%to the covering space
%$\Sigma$ is locally given by $z\approx b t^n$
%then on $\Sigma$ we have just an exponential  inserted at $t=0$:

Notice that $\sigma^P_n$ represents the twist operator corresponding 
to the specific permutation of indices (for example, $(1,2,\dots,n)$). This 
object is not well defined in the orbifold CFT, so to construct a proper operator one should sum over the conjugacy class \cite{DVVV89}:
\be\label{OrbTwist}
O^P_n=\left[\frac{1}{n(N-n)!N!}\right]^{1/2}\sum_{h\in G}
\sigma^P_{h(1\dots n)h^{-1}} 
\ee
In this paper we will mostly work with 
$\sigma^P_n$, and we will comment on going to $O^p_n$ in the end.    

\subsection{Correlators involving twist operators}

We will be interested in correlation functions involving two twist operators, so we begin with outlining a general procedure for evaluating correlators containing two twists and an arbitrary number of single--valued fields ${\tilde A}_k(z)$:\footnote{The invariance of the correlation function under the action of $S_N$ implies that the twist operators must have the same order $n$.}
\bea\label{TwoPntTwst}
\langle \sigma^P_n(0)\sigma^{Q\dagger}_n(a) 
{\tilde A}_1(z_1)\dots {\tilde A}_m(z_m)\rangle
\eea
The goal of this subsection is to write such correlators in
a form which does not contain twists, this will be accomplished by passing to the covering space of the $z$ plane. 

To define a chiral operator $\sigma_n^P(0)$,  we start by cutting a hole of radius $\epsilon$ in the $z$
plane around $z=0$ \cite{lm,lmtwo}. As we circle around this hole, $n$ different copies of
the $c=6$ CFT permute into each other.
The chiral operator with $h=j_3$ is obtained by applying currents to the basic twist, this corresponds to the insertion  of (\ref{DefCoverTwist}) on the covering space.
The resulting operator
$\sigma_n^{\epsilon P}(0)$  in the $z$ space still depends on
the cutoff
$\epsilon$, and the normalized chiral operator is defined by
\be
\sigma^P_n(0)=\frac{\sigma^{\eps P}_n(0)}{\langle \sigma^{\eps P}_n(0)
\left\{\sigma^{\eps P}_n(1)\right\}^\dagger\rangle^{1/2}}.
\ee

Now consider a  correlator that contains two such chiral operators
with twists of order $n$, and also a set of
operators ${\tilde A}_1(z_1), \dots, {\tilde A}_m(z_m)$, which do not generate any twists. Since ${\tilde A}_k$ are operators in the orbifold
theory, they are symmetric under the
interchange of the copies of the
$c=6$ CFT. Such symmetric operators can be made, for example, by starting with an operator in the $c=6$ CFT and
taking the product of $N$ copies from each CFT, and an important example of such construction will be discussed in the next subsection (see equation 
(\ref{defSpecSpin})). The operators ${\tilde A}_k$
could also be made by taking a sum over
identical operators from each copy of the CFT, or by a combining products
and sums. The arguments below apply
to all choices of the operators ${\tilde A}_k$, but for concreteness we
assume that the operator is a product of
(identical) operators from each copy of the CFT:
\be
{\tilde A}_k(z_k)=\prod_{i=1}^N A_k^{(i)}(z_k)
\label{mone}
\ee

Let the first  $n$ copies of the $c=6$ CFT be permuted by the twist
operators, then the remaining $N-n$ copies give the
factor 
\be
\prod_{i=n+1}^N\langle A_1^{(i)}(z_1)\dots  A_m^{(i)}(z_m)\rangle
\label{mtwo}
\ee
in the correlation function (\ref{TwoPntTwst}). 
The contribution from the first $n$ copies of the $c=6$ CFT to  (\ref{TwoPntTwst}) is
\be\label{defEqn2}
\langle \sigma^P_n(0)\sigma^{Q\dagger}_n(a) A_1(z_1)\dots A_m(z_m)\rangle=
\frac{\langle \sigma^{\eps P}_n(0)\sigma^{\eps Q\dagger}_n(a) A_1(z_1)
\dots A_m(z_m)\rangle}{\langle \sigma^{\eps P}_n(0)
\left\{\sigma^{\eps P}_n(1)\right\}^\dagger\rangle^{1/2}
\langle \sigma^{\eps Q}_n(0)
\left\{\sigma^{\eps Q}_n(1)\right\}^\dagger\rangle^{1/2}},
\ee
where we defined
\be
A_k\equiv\prod_{i=1}^n A_k^{(i)}.
\ee

The numerator of the rhs of (\ref{defEqn2}) can be evaluated by going to the covering space $\Sigma$ of the $z$ plane, where the twists are resolved. We will denote the holomorphic coordinate on $\Sigma$ by $t$, the lift of the operators $A_k^{(i)}$ to the covering space by $A^t_k$, and $n$ images of the point $z_k$ by $t_{k,j}, j=1,\dots n$.
%$A_k^{(i)}$ in the CFT on the $z$ plane.
%Let $A^t_k$ be the
%same operator in the $c=6$ CFT on  $\Sigma$ as each of the operators
%$A_k^{(i)}$ in the CFT on the $z$ plane.  
%Further, let $t_{k,j}, j=1,
%\dots n$ be the $n$ images on $\Sigma$ of the
%point $z_k$ on the $z$ plane. 
The operator insertion $A_k(z_k)$ in
the $z$ plane corresponds to a product of
operators in the $c=6$ CFT on  $\Sigma$:
\be
    A_k(z_k)\rightarrow \prod_{j=1\dots n}
\left(\frac{dz}{dt}(t_{k,j})\right)^{-\Delta_k}A_{k}^t(t_{k,j})
\label{mthree}
\ee
Here $\Delta_k$ is the dimension of each of the $A_k^{(i)}$. To rewrite equation (\ref{defEqn2}) in terms of the covering space, we recall the relation (\ref{DefCoverTwist}).
The numerator of the rhs of (\ref{defEqn2}) becomes
\be\label{Sep5}
\langle \sigma^{\eps P}_n(0)\sigma^{\eps Q\dagger}_n(a) A_1(z_1)
\dots A_m(z_m)\rangle=
e^{S_L}
\prod_{k,j}\left(\frac{dz}{dt}(t_{k,j})\right)^{-\Delta_k}
\langle {\hat\sigma}^{P}_n(0)
{\hat\sigma}^{Q\dagger}_n(a) \prod A_k^t(t_{k,j})\rangle.
\ee
Here $e^{S_L}$ is a contribution of the conformal anomaly \cite{friedan}; in the present case 
it can be written as \cite{lm}
\be
e^{S_L}=a^{(n-1/n)/2}f(n,\eps),
\label{msev}
\ee
and it does not depend upon $P$ and $Q$.  

The
denominator of the rhs of
(\ref{defEqn2}) is
\be
\langle \sigma^{\eps P}_n(0)
\left\{\sigma^{\eps P}_n(1)\right\}^\dagger\rangle=f(n,\eps)\langle
\hat \sigma_n^P (0)\hat
\sigma_n^{P\dagger}(1)\rangle=f(n,\eps),
\label{msix}
\ee
where in the first step we have used (\ref{msev}) and in the second
step we have used normalization of twist operators.
Substituting (\ref{Sep5}) and (\ref{msix})  into  (\ref{defEqn2})
we get:
\bea\label{defEqn3}
&&\langle \sigma^P_n(0)\sigma^{Q\dagger}_n(a) A_1(z_1)\dots A_m(z_m)\rangle
\nonumber\\
&&\qquad\qquad =a^{-\frac{1}{2}(n-\frac{1}{n})}
\prod_{k,j}\left(\frac{dz}{dt}(t_{k,j})\right)^{-\Delta_k}
\langle {\hat\sigma}^{P}_n(0)
{\hat\sigma}^{Q\dagger}_n(a) \prod A_k^t(t_{k,j})\rangle.
\eea

To summarize, we have demonstrated the following fact, which is intuitively obvious. If there are only
two twist operators in the correlator then the contribution of the
conformal anomaly cancels out, and the
problem reduces to evaluation of correlation functions in the
$c=6$ CFT on the covering space $\Sigma$.
This happens because the chiral operators containing the
twists are normalized by means of their two-point
functions. By contrast, if we are computing three or four-point
functions as in \cite{lm, lmtwo}, then the
conformal anomaly gives a nontrivial contribution.

Finally we recall that the proper twist operator in the orbifold CFT is $O^P_n$ 
given by (\ref{OrbTwist}) rather then $\sigma^P_n$. This means that the actual 
correlator to be computed is 
\be
\langle O^P_n(0)O^{Q\dagger}_n(a) A_1(z_1)\dots A_m(z_m)\rangle
\ee
rather than (\ref{defEqn3}). Assuming that operators 
${\tilde A}_1(z_1),\dots,{\tilde A}_m(z_m)$ do not 
contain twists, we get:
\bea
&&\langle O^P_n(0)O^{Q\dagger}_n(a) {\tilde A}_1(z_1)\dots {\tilde A}_m(z_m)\rangle\nonumber\\
&&\qquad=
\frac{1}{n(N-n)!N!}\sum_{g\in G}\sum_{h\in G}
\langle \sigma^P_{g(1\dots n)g^{-1}}(0)
\sigma^{Q\dagger}_{h(1\dots n)h^{-1}}(a) 
{\tilde A}_1(z_1)\dots {\tilde A}_m(z_m)\rangle\nonumber\\
&&\qquad=
\frac{1}{n(N-n)!}\sum_{h\in G}
\langle \sigma^P_{(1\dots n)}(0)
\sigma^{Q\dagger}_{h(n\dots 1)h^{-1}}(a) {\tilde A}_1(z_1)\dots {\tilde A}_m(z_m)\rangle.
\eea
Note that the prefactor in front of the sum is equal to the number of non-vanishing terms in the sum, so we finally get
\be
\langle O^P_n(0)O^{Q\dagger}_n(a) {\tilde A}_1(z_1)\dots {\tilde A}_m(z_m)\rangle=
\langle \sigma^P_n(0)\sigma^{Q\dagger}_n(a) 
{\tilde A}_1(z_1)\dots {\tilde A}_m(z_m)\rangle.
\ee
We will use this formula to evaluate correlation functions involving spin operators, the objects which are constructed in the next subsection.   

\subsection{Spin operators}

An $N=4$ CFT has the unique vacuum $|0\rangle_{NS}$ with $j=h=0$, which belongs to the Neveu-Schwarz (NS) sector.  However, the black hole of interest arises in the Ramond (R) sector, where there are $e^{2\sqrt{2}\sqrt{N}}$ degenerate ground states with 
$h={c\over 24}$. Such states can be created by 
acting on  $|0\rangle_{NS}$ by a spin operator ${\cal S}$:
\be
|0\rangle_R = {\cal S} |0\rangle_{NS}\,.
\label{aaone}
\ee
The states $|i\rangle$ and $|f\rangle$ in Fig.1 are the R ground states  of the D1-D5 system. We will choose  $|i\rangle, |f\rangle$ from a 
special subset
of R ground states, which we now describe.

All Ramond ground states can be obtained by starting with chiral primaries in the NS sector and applying the operation of spectral flow \cite{SpFlowSS}, which maps the state with charges 
$(h,j_3)$ in the NS sector into state with charges $(h',j_3')$ in the R sector:
\bea
h'&=&h-j_3+{c\over 24},\nonumber\\
j_3'&=&j_3-{c\over 12}.
\label{lone}
\eea
A general chiral primary of our CFT is given by a twist operator \cite{lmtwo}, then the corresponding R vacuum contains twist as well, and such states are discussed in section \ref{SubsInterm}. Here we will describe a subset of 
Ramond vacua which do not interchange different copies of $c=6$ CFT, such states are obtained by applying the spectral flow to the NS vacuum.  

For one copy of the $c=6$ CFT the spectral flow of $|0\rangle_{NS}$ gives a R ground state with $h={1\over 2}$ 
and SU(2) quantum numbers
$j={1\over 2}, m=-{1\over 2}$. In the free CFT with
bosonized fermions
we find the unique vertex operator creating the state with these
properties\footnote{We recall that the first four fields 
$\phi_1,\phi_2,\phi_3,\phi_4$ come from the original scalars, so they are not affected by the spectral flow. To modify the boundary conditions for the fermions, one has to insert exponential factors containing the bosonized fields $\phi_5$ and $\phi_6$, this explains appearance of $e_a$ in (\ref{Sep08}).}:
\be\label{Sep08}
S_-(z)=:\exp\left(-\frac{i}{2}e_a\phi_a(z)\right):
\ee
Since this operator has an $SU(2)$ spin $j=1/2$, it is a member of a doublet;
     the other component of the doublet can be obtained by applying
$J_0^+$ to $S_-(z)$:
\be
S_+(z)=:\exp\left(\frac{i}{2}e_a\phi_a(z)\right):
\ee

Now consider $N$ copies of ${\cal N}=4$ CFT with total central charge
$c=6N$. The NS  vacuum $|0\rangle_{NS}$
flows to a state with $h=N/4$, $j_3=-N/2$, which is a member of an SU(2)
multiplet with
$j=N/2$.  For the top member of this multiplet ($j=m={N\over 2}$) the
appropriate  vertex operator is the product of the spin field
$S_+(z)$ from all the different copies of the $c=6$
CFT:
\be\label{defSpecSpin}
S_{N,0}(z)\equiv S_{+\dots +}(z)=\prod_{i=1}^N:\exp\left(\frac{i}{2}e_a
\phi_i^a(z)\right):
\ee
The other elements of the SU(2) representation  can be constructed
using $J^-_0$:
\be
\label{defGenerSpin}
S_{N-k,k}(z)\equiv\sqrt{\frac{(N-k)!}{k!N!}}\left(J^-_0\right)^k 
S_{N,0}(z),\qquad
k=0,\dots N.
\ee
These operators are normalized by
\be\label{SpinNormal}
\langle S_{N-k,k}(0)S^\dagger_{N-k,k}(z)\rangle =z^{-N/2}.
\ee
We will be interested in processes for which the states $|i\rangle, |f\rangle$ in Fig.1 have the form  
(\ref{aaone}) with ${\cal S}$ given by (\ref{defGenerSpin}).

Recalling the relation (\ref{defSpecSpin}), we see that  the operator
$S_{N-k,k}$ generates a linear combination
of states where each copy of the $c=6$ CFT  is in one of the two
$j=1/2$  Ramond ground states:  $N-k$ of these
copies have the state with spin pointing up ($j_3=1/2$) and $k$ of
these copies have the spin pointing
down ($j_3=-1/2$). The linear combination (\ref{defGenerSpin}) produces the states which are symmetric under the
interchange of copies, as is required of an operator in the orbifold
theory $M^N/S^N$.

%==============================================================
%                   Cylinder figure
%==============================================================
\begin{figure}
\hskip 2.5in
\begin{tabular}{rl}
\begin{picture}(85.00,15.00)
\put(0,10){$S^\dagger_{N-l,l}$}
\end{picture}&\\
\epsfysize=3in \epsffile{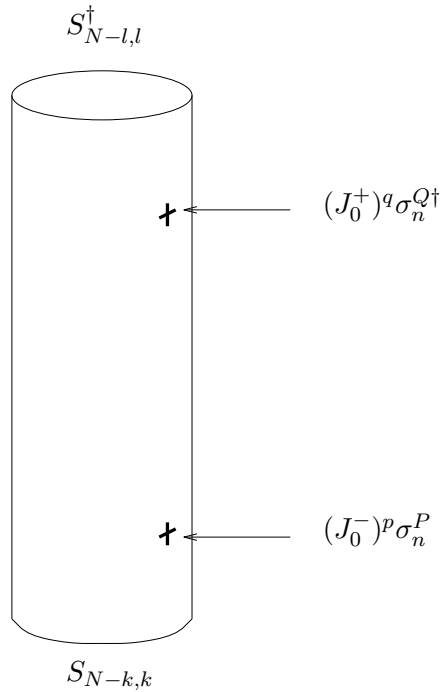}&
\begin{picture}(70.00,80.00)
\put(0,40){$(J^-_0)^p\sigma_n^P$}
\put(0,163){$(J^+_0)^q\sigma_n^{Q\dagger}$}
\end{picture}\\
\begin{picture}(85.00,15.00)
\put(0,5){$S_{N-k,k}$}
\end{picture}&
\end{tabular}
\caption{\label{FigCyl} A more explicit version of figure 1(b): the microstates of a black hole correspond to spin operators 
$S_{N-k,k}$ and $S^\dagger_{N-l,l}$, absorbed and emitted quanta correspond chiral primaries represented by (rotated) twist operators.}
\end{figure}

\section{S--matrix element and correlation function in the CFT}
\label{SectSmatr}
\renewcommand{\theequation}{3.\arabic{equation}}
\setcounter{equation}{0}

After defining the spin and twist operators in the previous section, we can construct the correlation function which models  
absorption/emission of quanta by a black hole. On the gravity side we consider a process depicted in figure 1(a), where the initial and final states, $|i\rangle$ and $|f\rangle$, are described by regular geometries corresponding to the simplest Ramond vacua \cite{mm,LMmetr}. 

The processes associated with black hole physics are encoded in the CFT defined on a cylinder.  We denote the
compact spatial direction of this cylinder by $\sigma$ ($0\le \sigma
< 2\pi$), and the orthogonal noncompact
direction by\footnote{We work with Euclidean signature on this cylinder and rotate
final results to Lorentzian signature at
the end, as is usual for CFT computations.} $\tau$. The cylinder can be mapped to a plane by a conformal transformation, so 
most of our work will be done on the plane, which is parameterized by the
coordinate $z$, and the map to the cylinder will be discussed 
at the end. We will assume that the endpoints of the cylinder ($\tau=\pm\infty$) map to points $z=v$ and $z=w$ in the plane. There will be four operator insertions in this plane
(their pictorial representation is shown in Figure \ref{FigCyl}):

\begin{enumerate}[(a)]
\item{The first R vacuum with $j=N/2$ is introduced at
$\tau\rightarrow -\infty$ on the cylinder. If $\tau=-\infty$ maps to $z=v$ in the plane, then the required  state is created by
inserting an operator $S_{N-l,l}(v)$ at the point $z=v$ in the plane. In gravity, this state corresponds to the metric of \cite{mm,LMmetr} rotated by $l$ units.}

\item{The second R ground state with
$j=N/2$ is introduced at $\tau\rightarrow\infty$ on the cylinder,
which maps to $z=w$ . We thus insert an operator
$S_{N-k,k}^\dagger(w)$ at this point. In gravity, this state corresponds to the metric of \cite{mm,LMmetr} rotated by $k$ units.}

\item{We are interested in the process describing an excitation of the ground state by a supergravity quantum. In the CFT this particle is represented by (an SU(2) rotation of) a
chiral operator $O_n^P$, which is written as a linear combination of operators $\sigma_n^P$. 
We will begin with computing the correlator for an insertion of
$\sigma_n^P$ and sum over permutations later. Using translational
invariance in the $z$ plane, we put this
operator at
$z=0$.}

\item{The system de-excites to the final Ramond ground state by
emitting another supergravity
quantum, which is represented in the CFT by (an SU(2) rotation of) an operator
$(O_n^Q)^\dagger$. Again, we begin with computing the
correlator for a given
$\sigma_n^{Q\dagger}$. Let
$\sigma_n^{Q\dagger}$ be inserted at $z=a$.
Since $\sigma_n^P, \sigma_n^{Q\dagger}$ are the only twist operators
appearing in the amplitude, their twist orders
must be equal, and we have taken each to be $n$. The SU(2) charges of all
the four operators in the correlator must
add up to zero.}
\end{enumerate}

\noindent
To summarize, we are interested in evaluating the correlator
\be\label{def4pointB}\label{genSSss1}
\langle S_{N-l,l}(v)S^\dagger_{N-k,k}(w)
\left[(J_0^-)^p\sigma^{P}_n\right](0)
\left[(J_0^+)^q\sigma^{Q\dagger}_n\right](a)\rangle_{S_N},
\ee
and the relevant calculations will be performed in the next few sections. The superscript $S_N$ in (\ref{def4pointB}) is introduced to stress the fact that this correlation function is computed in $M^N/S_N$ orbifold. Without loss of generality, we will set $l=0$ in (\ref{genSSss1}) by performing a global SU(2) rotation.

\section{Outline of the calculation}
\label{SectGen4ptRed}
\renewcommand{\theequation}{4.\arabic{equation}}
\setcounter{equation}{0}

Computation of all four--point functions (\ref{def4pointB}) on $M^N/S_N$ orbifold can be reduced to evaluation of three simple correlators in $M^n/S_n$ theory, and in this section we will outline the logic behind this reduction. The details of computations are presented in the appendix \ref{AppRdct}.

The reduction to the basic correlators is performed in three steps.
\begin{enumerate}
\item{Using the properties of spin operators, one can reduce the four--point function on $M^N/S_N$ orbifold to its counterpart in 
$M^n/S_n$ theory (see appendix \ref{AppRdct}):
\bea\label{TempLargeN}
&&\langle S_{N,0}(v)S^\dagger_{N-k,k}(w)\left((J_0^-)^p\sigma^P_n\right)(0)
\left((J_0^+)^q\sigma^{Q\dagger}_n\right)(a)\rangle_{S_N}
%\nonumber\\
%&&\qquad
=(v-w)^{-(N-n)/2}\nonumber\\
&&\quad\times \left(\frac{(N-k)!n!}{N!(n-k)!}\right)^\frac{1}{2}\langle
{S}_{n,0}(v){S}^{\dagger}_{n-k,k}(w)
\left((J_0^-)^p\sigma^P_n\right)(0)
\left((J_0^+)^q\sigma^{Q\dagger}_n\right)(a)\rangle_{S_n}.
\eea
This reduces the problem to evaluation of the correlation functions
\bea\label{May26}
\langle
{S}_{n,0}(v){S}^{\dagger}_{n-k,k}(w)
\left((J_0^-)^p\sigma^P_n\right)(0)
\left((J_0^+)^q\sigma^{Q\dagger}_n\right)(a)\rangle_{S_n}.
\eea
}
\item{To simplify the correlators (\ref{May26}), one can move 
operators $J_0^-$ from
$\sigma^P_n$ to the other insertions in the correlator, this corresponds to a global $SU(2)$ rotation. Thus
(\ref{May26}) can be written as a weighted sum over the following
set of correlation functions for different values of $k,l$:
\be\label{genSSss}
\langle S_{n-l,l}(v)S^\dagger_{n-k,k}(w)\sigma^P_n(0)
\left((J_0^+)^s\sigma^{Q\dagger}_n\right)(a)\rangle_{S_n}.
\ee
 }
\item{Moving around $J_0^+$ and using the fact that 
$J_0^+\sigma^P_n(0)=0$, one can rewrite (\ref{genSSss}) in terms of 
``basic" 4--point functions (see appendix \ref{AppRdct}):
\bea\label{SumRule}
&&\langle S_{n-l,l}(v)S^\dagger_{n-k,k}(w)\sigma^P_n(0)
\left((J_0^+)^s\sigma^{Q\dagger}_n\right)(a)\rangle_{S_n}\nonumber\\
&&\qquad=\sum_{p=0}^s \frac{s!}{p!(s-p)!}
\left(\frac{(n-k+p)!k!}{(n-k)!(k-p)!}
\frac{(n-l)!(s-p+l)!}{(n-l-s+p)!~l!}\right)^{1/2}\\
&&\qquad\quad\times
\langle S_{n-l-s+p,l+s-p}(v)
S^\dagger_{n-k+p,k-p}(w)
\sigma^P_n(0)\sigma^{Q\dagger}_n(a)\rangle_{S_n}.\nonumber
\eea
}
\end{enumerate}

These steps reduce evaluation of the general 4--point function (\ref{def4pointB}) to computation of the basic 
correlators\footnote{Here and below we drop the superscript $S_n$: the rank of the orbifold can be read off from the order of spin operators.}
\bea
\langle S_{n-l,l}(v)
S^\dagger_{n-k,k}(w)
\sigma^P_n(0)\sigma^{Q\dagger}_n(a)\rangle
\eea
on $M^n/S_n$ orbifold.

Recalling the construction of chiral operators in section \ref{SectTwist}, we find 
four possible values for the charges $P, Q$:
\bea
\label{PQpairSame}
(P,Q)=\left(\frac{n-1}{2},\frac{n-1}{2}\right),\qquad
(P,Q)=\left(\frac{n+1}{2},\frac{n+1}{2}\right),\\
\label{PQpairDiff}
(P,Q)=\left(\frac{n+1}{2},\frac{n-1}{2}\right),\qquad
(P,Q)=\left(\frac{n-1}{2},\frac{n+1}{2}\right).
\eea
This reduces evaluation of (\ref{def4pointB}) to calculation of two
basic 4--point functions:
\bea\label{BasicCorr1}
\mbox{(a)}&&\langle S_{n-k,k}(v)S^\dagger_{n-k,k}(w)
\sigma^Q_n(w)\sigma^{Q\dagger}_n(a)\rangle, ~~~Q={n\over 2}, {n+1\over 2}
\eea
corresponding to the charges (\ref{PQpairSame}), and
\bea\label{BasicCorr2}
\mbox{(b)}&&\langle S_{n-k+1,k-1}(v)S^\dagger_{n-k,k}(w)\sigma^{\frac{n+1}{2}}_n(0)
\sigma^{-\frac{n-1}{2}}_n(a)\rangle\qquad\qquad
\eea
corresponding to the first set of charges in (\ref{PQpairDiff}). Note
that one does
not have to consider the second set in (\ref{PQpairDiff}) separately, since
\bea
&&\langle S_{n-k,k}(v)
S^\dagger_{n-k-1,k+1}(w)\sigma^{\frac{n-1}{2}}_n(0)
\sigma^{-\frac{n+1}{2}}_n(a)\rangle\nonumber\\
&&\qquad=\langle S^\dagger_{n-k,k}(v)S_{n-k-1,k+1}(w)
\sigma^{-\frac{n-1}{2}}_n(0)
\sigma^{\frac{n+1}{2}}_n(a)\rangle.
\eea

To summarize, in order to calculate the general four point function
(\ref{genSSss}), one has to evaluate two basic correlators
(defined by (\ref{BasicCorr1}) and (\ref{BasicCorr2})), and use the
rule (\ref{SumRule}).
The basic correlators will be evaluated in the section \ref{SectBasic}, and the complete
expression for the general four point function is derived in the Appendix
\ref{SectGenCorr}. The basic correlator can be reduced to a meromorphic function of one complex variable, and 
in the next section we will discuss the behavior of this function near its poles and its physical interpretation. These results will be used in section \ref{SectBasic} to recover the correlators.

\section{Basic correlators and three--point functions}
\renewcommand{\theequation}{5.\arabic{equation}}
\setcounter{equation}{0}

To evaluate the basic correlators (\ref{BasicCorr1}) and (\ref{BasicCorr2}), we begin with reducing them to functions of one complex variable
and studying analytic properties of such functions. Conformal invariance determines (\ref{BasicCorr1}) and (\ref{BasicCorr2}) up to a function of one complex cross ratio $x$,
\be\label{DefCrRat}
x=\frac{v(w-a)}{w(v-a)}. 
\ee
As shown in the Appendix \ref{AppA}, the relevant $x$--dependent combination is
\bea\label{DefFIndFin}
&&f(n,l,k|P,p;Q,q|x)\equiv
\langle S_{n-l,l}(v)S_{n-k,k}^\dagger(w)
\left[(J_0^-)^p\sigma^{P}_n\right](0)
\left[(J_0^+)^q\sigma^{Q\dagger}_n\right](a)\rangle\nonumber\\
&&\phantom{\frac{e^x}{f^y}}\times
v^{-(Q-2P-n/4)/3}w^{-(Q-2P-n/4)/3}(w-a)^{(2Q-P+n/4)/3}\nonumber\\
&&\phantom{\frac{e^x}{f^y}}\times
(-a)^{(2P+2Q-n/2)/3}(v-w)^{(n-P-Q)/3}(v-a)^{(2Q-P+n/4)/3}
\eea
Here the first three parameters 
of $f$ appear in  the spin operators, the next four describe
charges of the twist operators, and the last parameter is the cross ratio.

Function $f(x)$ depends on one complex
variable, it has 
singularities only at $x=0, 1, \infty$; furthermore, these singularities
are poles. Thus we can
determine $f$ by finding the orders and residues for these poles. It
turns out that we will only need the residue of the leading pole at each singular
point -- this fact
leads to a great simplification in the calculation.
To find the orders and residues for the poles, we will examine the
OPE where two of
the insertions in the 4-point function approach each other, 
identify  the leading operator appearing in this OPE, and evaluate the corresponding fusion coefficient.

\subsection{Four point function and operator product expansions.}

We begin with expressing the correlator
\be\label{BasicCorr1p}
\langle S_{n-k,k}(v)S^\dagger_{n-k,k}(w)
\sigma^Q_n(0)\sigma^{Q\dagger}_n(a)\rangle.
\ee
in terms of fusion coefficients.  It is  convenient to place one
operator at infinity, in which
case we must consider the ratio
\be
\lim_{a\rightarrow\infty}\frac{\langle S_{n-k,k}(v)S^\dagger_{n-k,k}(w)
\sigma^Q_n(0)\sigma^{Q\dagger}_n(a)\rangle}{
\langle \sigma^Q_n(0)\sigma^{Q\dagger}_n(a)\rangle}
\label{zone}
\ee

To establish notation we recall the following elementary steps valid
for any CFT.
Suppose one wants to evaluate the ratio:
\be\label{ABCDrat1}
\frac{\langle A(w)B(v)C(0)D^\dagger(\infty)\rangle}{
\langle D(0)D^\dagger(\infty)\rangle}
\ee
Consider the OPE
\be\label{ABCDrat2}
B(v)C(0)=\sum_i C_{B,C,i}v^{-\Delta_B-\Delta_C+\Delta_i}{\cal A}_i(0),
\ee
where $\Delta_i$ is a conformal dimension of ${\cal A}_i$. We
assume that the set of operators appearing on the rhs of
(\ref{ABCDrat2}) is orthonormal:
\be\label{OiNorm}
\langle {\cal A}_i(0){\cal A}^\dagger_j(1)\rangle =\delta_{ij}.
\ee
Then substituting the expansion (\ref{ABCDrat2}) and an analogous expansion 
for $A(z)D^\dagger(0)$ into (\ref{ABCDrat1}), we get:
\bea\label{ABCDrat3}
\frac{\langle A(w)B(v)C(0)D^\dagger(\infty)\rangle}{
\langle D(0)D^\dagger(\infty)\rangle}&=&
\sum_i C_{B,C,i}v^{-\Delta_B-\Delta_C+\Delta_i}\frac{\langle A(w){\cal A}_i(0)
D^\dagger(\infty)\rangle}{
\langle D(0)D^\dagger(\infty)\rangle}\nonumber\\
&=&\sum_i C_{B,C,i}\left(C_{A^\dagger,D,i}\right)^*
v^{-\Delta_B-\Delta_C+\Delta_i}
w^{-\Delta_i-\Delta_A+\Delta_D}
\eea
In particular, if one is interested in the limit of $v/w\ll 1$, then the
leading contribution arises from the operator ${\cal A}_i$ with the lowest
possible dimension. In the next subsection we will evaluate such a leading
contribution for the OPE involving a spin operator and a twist operator.

%==============================================================
%                   Combined 3pt figure
%==============================================================
\begin{figure}
\begin{tabular}{ccc}
\begin{tabular}{c}
\epsfysize=1.5in \epsffile{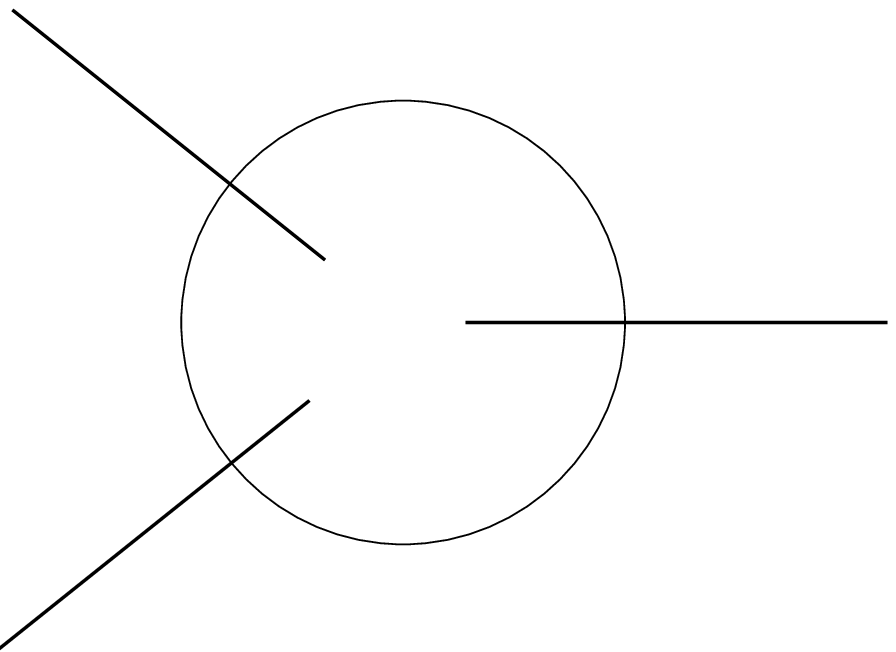}\\
\begin{picture}(150.00,10.00)
\put(-10,3){$S_{k,n-k}$}
\put(-10,130){$\sigma_n^{\frac{n+1}{2}}$}
\put(160,70){${\cal A}_{n,k}$}
\end{picture}
\end{tabular}&
\begin{picture}(50.00,40.00)
\end{picture}&
\begin{tabular}{ccc}
\epsfysize=1.2in \epsffile{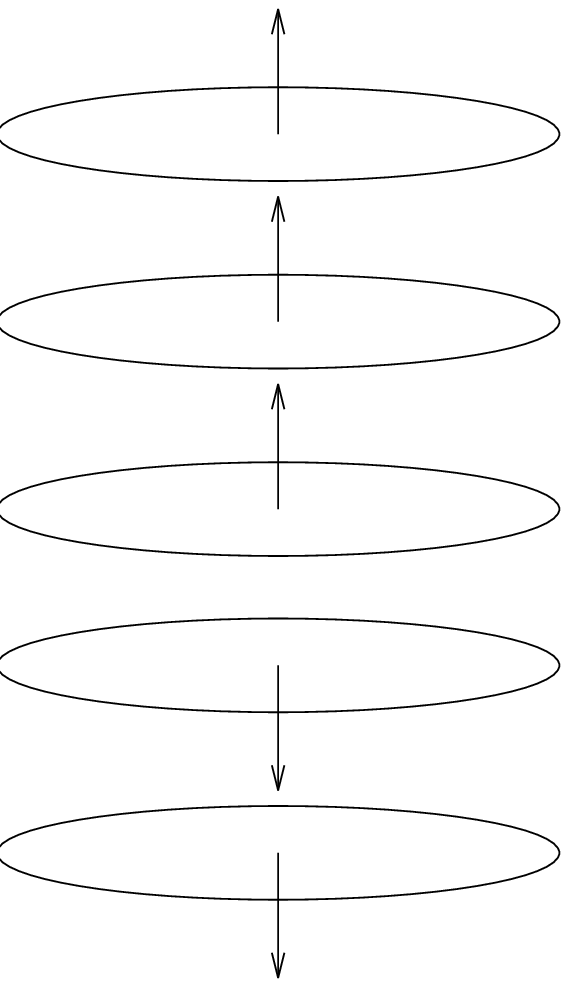}&
\begin{picture}(30.00,90.00)
\put(-10,40){\vector(1,0){40}}
\put(0,50){$\sigma_n^{\frac{n+1}{2}}$}
\end{picture}&
\epsfysize=1.2in \epsffile{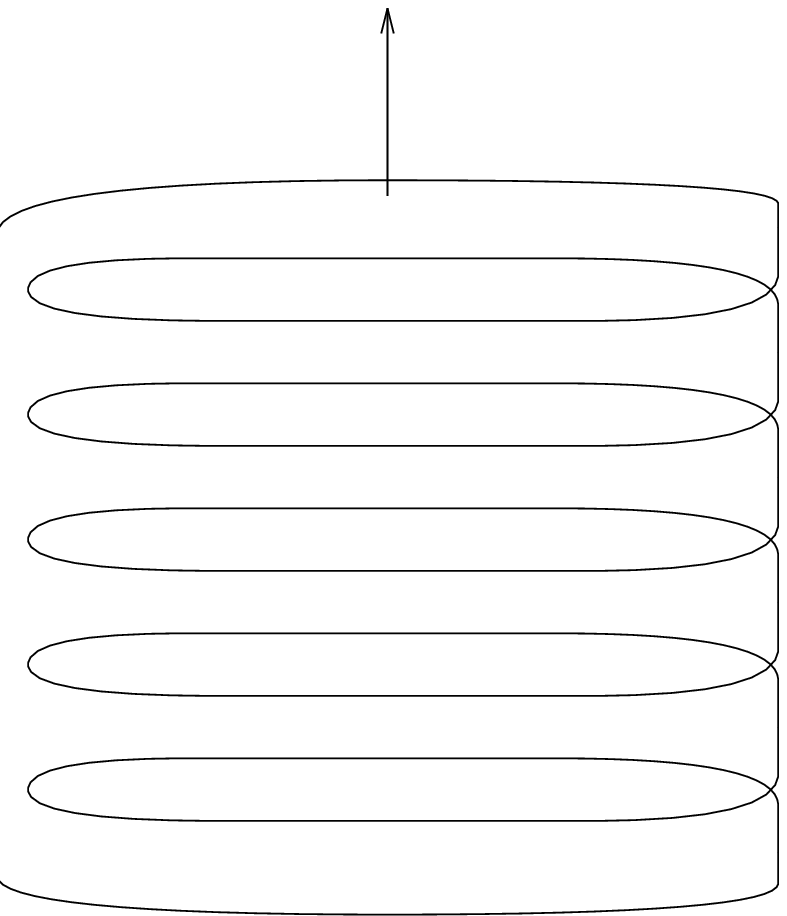}\\
\begin{picture}(70.00,30.00)
\put(25,10){$S_{k,n-k}$}
\end{picture}&&
\begin{picture}(70.00,30.00)
\put(35,10){${\cal A}_{n,k}$}
\end{picture}
\end{tabular}\\
(a)&&(b)
\end{tabular}
\caption{\label{fig3pt} (a) Three--point function involving the `intermediate state' in the CFT.\newline
(b) Interpretation of the same process in terms of multiwound string.
}
\end{figure}
%==============================================================

\subsection{OPE of twist operator with spin operator.}
\label{SectIntOPE}
As shown in the Appendix \ref{AppThreePoint}, the leading contributions to
the OPEs of twist operators and spin operators are given by (see
(\ref{Final-OPE}), (\ref{Final+OPE})):
\bea\label{SpinTwistOPE}
S_{k,n-k}(w)\sigma^{\frac{n+1}{2}}_n(0)&=&C(n,k,\frac{n+1}{2})
w^{-\frac{n+1}{2}+k}{\cal A}_{n,k}(0)+
O(w^{-\frac{n+1}{2}+k+1}),\\
S_{k,n-k}(w)\sigma^{\frac{n-1}{2}}_n(0)&=&C(n,k,\frac{n-1}{2})
w^{-\frac{n-1}{2}-\frac{n}{4}+(\frac{n}{4}-1+k)}{\cal A}_{n, k-1}(0)+
O(w^{-\frac{n-1}{2}+k})
\eea
Operators ${\cal A}_{n, k}$ in the rhs of these expressions are normalized,
\bea
\langle{\cal A}_{n, k}(0){\cal A}^\dagger_{n, k}(z)\rangle
=z^{-2\Delta_{n, k}},
\eea
their dimensions $\Delta_{n, k}$ and SU(2) quantum numbers $(j,j_3)$ are 
\bea
\Delta_{n, k}=\frac{n}{4}+k,\qquad j_3=j=k+\frac{1}{2}
\eea
The fusion coefficients are given by (\ref{Lead+C}) and (\ref{LaedC}),
\be\label{Fusion26}
C(n,k,\frac{n+1}{2})=\left(\frac{n-k}{n}\right)^{1/2},\qquad
C(n,k,\frac{n-1}{2})=-\left(\frac{k}{n}\right)^{1/2}.
\ee
We will also present the expression for the image of ${\cal A}_{n, k}$ in the covering
$t$ plane (see (\ref{mergeNN}), (\ref{OPE+3})):
\be
\hat{\cal A}_{n,k}(0)=
\left(\frac{(n-k-1)!}{k!(n-1)!}\right)^{1/2}
\left({\hat J}^+_{-n}\right)^k
\left[:\exp\left(\frac{i}{2}e_a\Phi^a(0)
\right):\right].
\ee
Using the explicit form of the map $z=t^n$ (see (\ref{ztMap})), the last
expression can be rewritten in terms of the operators in the
original $z$ plane:
\be\label{DefineCalA}
{\cal A}_{n,k}(0)=
\left(\frac{(n-k-1)!}{k!(n-1)!}\right)^{1/2}
\left(J^+_{-1}\right)^k {\cal A}_{n,0}(0)
\ee
Fusion coefficients (\ref{Fusion26}) supply all necessary information for evaluating the four--point function (\ref{BasicCorr1p}) using (\ref{ABCDrat3}),
and the relevant computation will be discussed in the next section. We conclude this section by describing the nature of the
operators ${\cal A}$ that arise from the OPE of spin
operators $S$ and chiral twist operators $\sigma$.

\subsection{The operators corresponding to
`intermediate states' }
\label{SubsInterm}
 
\begin{figure}
\begin{tabular}{ccc}
\begin{tabular}{l}
\epsfysize=1.5in \epsffile{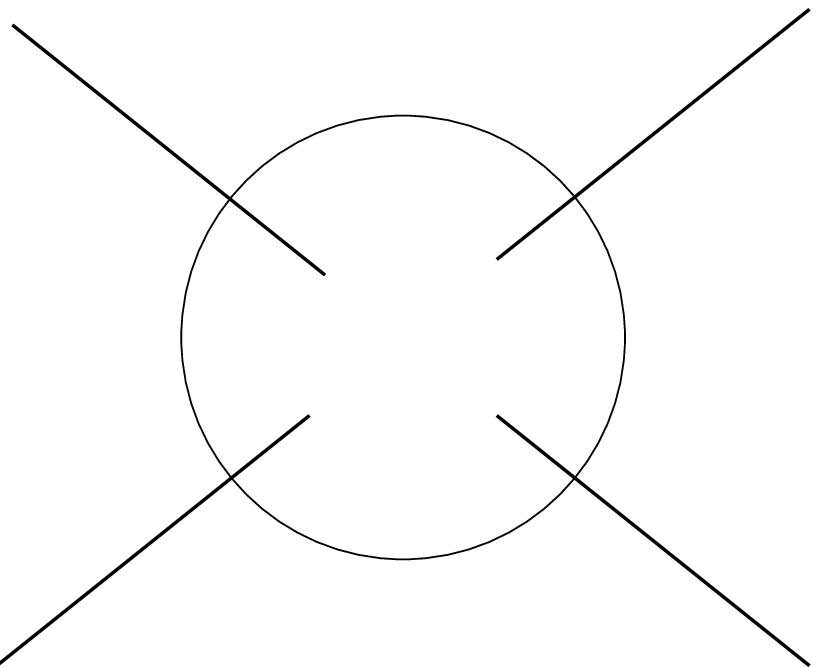}\\
\begin{picture}(150.00,10.00)
\put(-10,3){$S$}
\put(-10,125){$S^\dagger$}
\put(135,3){$\sigma$}
\put(135,125){$\sigma^\dagger$}
\end{picture}
\end{tabular}
&
\begin{picture}(50.00,40.00)
\end{picture}&
\begin{tabular}{l}
\epsfysize=0.62in \epsffile{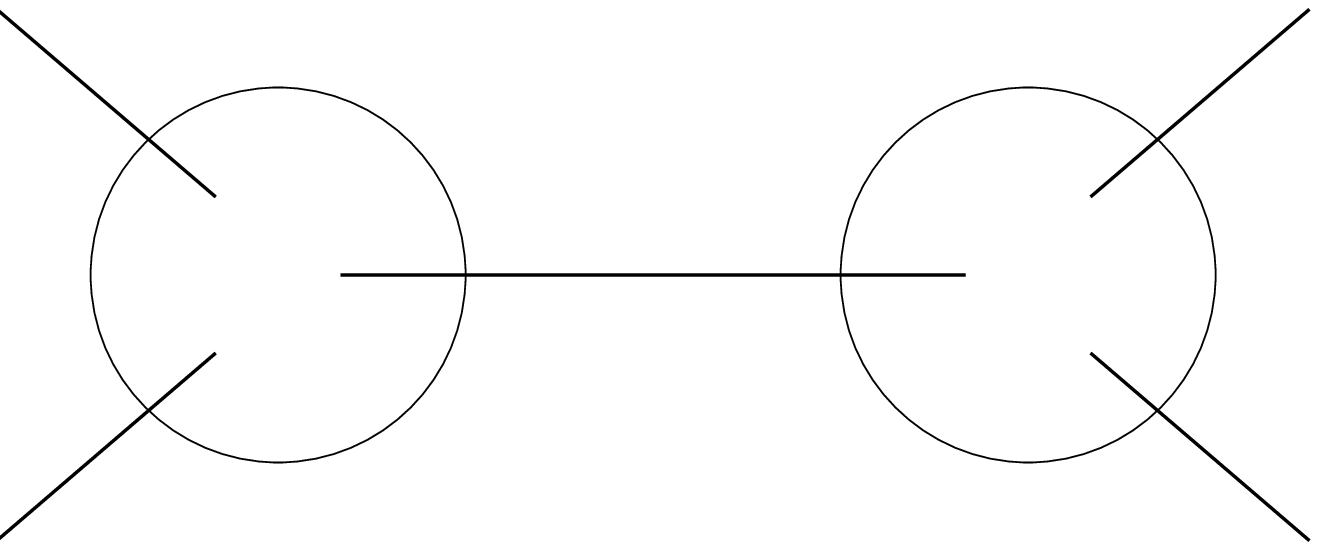}\\
\begin{picture}(200.00,10.00)
\put(-10,3){$S$}
\put(-10,65){$S^\dagger$}
\put(115,3){$\sigma$}
\put(115,65){$\sigma^\dagger$}
\end{picture}\\
\begin{picture}(100.00,30.00)
\end{picture}\\
\epsfysize=0.62in \epsffile{4to3.eps}\\
\begin{picture}(200.00,10.00)
\put(-10,3){$S$}
\put(-10,65){$\sigma^\dagger$}
\put(115,3){$\sigma$}
\put(115,65){$S^\dagger$}
\end{picture}\\
\begin{picture}(200.00,30.00)
\end{picture}\\
\epsfysize=0.62in \epsffile{4to3.eps}\\
\begin{picture}(200.00,10.00)
\put(-10,3){$S$}
\put(-10,65){$\sigma$}
\put(115,3){$\sigma^\dagger$}
\put(115,65){$S^\dagger$}
\end{picture}\\
\end{tabular}\\
(a)&&{\hskip -1in (b)}\\
\end{tabular}
\caption{\label{figFusion} 
(a) The 4---point function evaluated in this article.\newline
(b) The limits used to recover this function.
}
\end{figure}

Figure \ref{figFusion} shows that the 4-point
function (\ref{BasicCorr1p})  can be decomposed into 3-point functions
$S\sigma\rightarrow {\cal A}$, 
${\cal A}\rightarrow\sigma S$, where the operators ${\cal A}$ define
the `intermediate state' that travels up the
cylinder between the $\sigma$ insertions. As
mentioned above, the leading part of the OPE
$S\sigma\sim {\cal A}$ is sufficient to reconstruct
the 4-point function $\langle S\sigma \sigma S\rangle$, so the
operators ${\cal A}_{n,k}$ in eq. (\ref{DefineCalA}) are of central
interest in our analysis.

It is convenient to begin with considering the case $k=0$, i.e.,
to look at the OPE
\be
S_{0,n}(w)\sigma_n^{n+1\over 2}(0)\sim
w^{-{n+1\over 2}} {\cal A}_{n,0}(0).
\ee
The operator ${\cal A}_{n,0}$ has charge 
\be
j={1\over 2}
\label{mmone}
\ee
and dimension
\be
\Delta ={n\over 4}
\label{mmtwo}
\ee

The operator ${\cal A}_{n,0}$ possesses the `topological
characteristics' of both the spin field $S$ and the
chiral twist operator $\sigma$.  The operator
$\sigma_n^{n+1\over 2}$ joins $n$ copies of the
$c=6$ CFT on a unit to give one copy of the $c=6$ theory on
an circle of length $n$. The spin operator
$S_{0,n}$ acts within each copy of the $c=6$ CFT but
changes the boundary conditions of the fermions
from the NS  type to R type.  A little inspection
shows that {the operator
${\cal A}_{n,0}$ creates the R vacuum of the $c=6$ CFT
obtained by joining together $n$ copies of the $c=6$
CFT}.

This fact can be seen in more detail as follows.  We note that  
operator ${\cal A}_{n,0}$ has been constructed by passing to the $n$-fold cover of the $z$ 
space by the map $z\sim t^n$, and then inserting the operator $e^{\frac{i}{2}e_a\Phi^a}$.  
The twist, given by the map to the cover, tells us that the state created by ${\cal A}_{n,0}$ belongs to 
the $c=6$ CFT obtained by joining together $n$ copies of the $c=6$ CFT.  The vertex 
insertion makes the fermions in the $t$ space anti-periodic around the origin. The charge
$j=1/2$  of the vertex insertion is the just the charge of the R ground
state of a $c=6$ theory. The same argument implies that the general intermediate state represented by operator
\be
{\cal A}_{n,k}= \left(J_{-1}\right)^k {\cal A}_{n,0}
\ee
with $k>0$ creates an {\it excited state} in the Ramond sector ($J_{-1}$ raises the energy of the Ramond vacuum ${\cal A}_{n,0}$). We recall that ground states in the Ramond sector have a useful representation in terms of a multiwound string, in particular the vacuum ${\cal A}_{n,0}$
is represented by a string with winding $n$. This leads to a pictorial representation of the excited state ${\cal A}_{n,k}$ given in figure \ref{fig3pt}(b).

%==============================================================
%                   4 to 3 figure
%==============================================================

%==============================================================

\section{Evaluation of the basic correlators.}
\label{SectBasic}
\renewcommand{\theequation}{6.\arabic{equation}}
\setcounter{equation}{0}

Now we will use the properties of the intermediate state discussed in the previous section to characterize the singularities of the four point functions (\ref{BasicCorr1}), (\ref{BasicCorr2}), this will allow us to determine these correlators. For technical reasons, it is convenient to have a separate discussion of three branches, but there is no conceptual difference between these cases.  

\subsection{Evaluation of the basic correlator with $P=Q=\frac{n+1}{2}$.}

We begin with evaluation of the basic correlator (\ref{BasicCorr1}) for $Q=\frac{n+1}{2}$. It is convenient to start with the ratio
\be\label{++ratio}
{\cal F}(n,k,\frac{n+1}{2}|v,w)\equiv
\frac{\langle S_{n-k,k}(v)S_{n-k,k}^\dagger(w)
\sigma^\frac{n+1}{2}_n(0)\sigma^{\frac{n+1}{2}\dagger}_n(\infty)\rangle}
{\langle\sigma^\frac{n+1}{2}_n(0)\sigma^{\frac{n+1}{2}\dagger}_n(\infty)
\rangle}
\ee
and restore $a$--dependence using the general properties of CFT. Function (\ref{++ratio}) is analytic in the $v$ plane, and the only potential
singularities  occur if $v$ goes to $0$, $w$ or $\infty$. Since ${\cal F}$
is essentially a function of the cross ratio $x=v/w$ (apart from a
known overall factor),
the limit $v\rightarrow\infty$ is the same as
$w\rightarrow 0$. Thus the only possible singularities of the
holomorphic function ${\cal F}$ are located at $v=w$, $v=0$, $w=0$, and we now analyze the behavior of ${\cal F}$ near these points. 

\begin{enumerate}
\item{
In the limit $v\rightarrow w$, the spin operators
merge together, and at leading order they produce the NS vacuum:
\be
S_{n-k,k}(v)S^\dagger_{n-k,k}(w)=\frac{1}{(v-w)^{n/2}}+\dots.
\ee
This determines one of the asymptotics of ${\cal F}$:
\be\label{Asy+1}
{\cal F}(n,k,\frac{n+1}{2}|v,w)=\frac{1}{(v-w)^{n/2}}+
O\left(\frac{1}{(v-w)^{n/2-1}}\right).
\ee
}
\item{
In the limit $w\rightarrow 0$, the relation (\ref{ABCDrat3}) gives
\bea\label{Asy+2}
{\cal F}(n,k,\frac{n+1}{2}|w,v)&=&\sum_i \left|C_i(n,k,\frac{n+1}{2})\right|^2
w^{-\frac{n+1}{2}-\frac{n}{4}+\Delta_i}
v^{\frac{n+1}{2}-\frac{n}{4}-\Delta_i}\nonumber\\
&=&\frac{n-k}{n}w^{-\frac{n+1}{2}+k}
v^{\frac{1}{2}-k}+O(w^{-\frac{n+1}{2}+k+1})
\eea
For the leading order term in  $w$ we have used the dimension and fusion
coefficient  given by (\ref{Lead+Dim}) and (\ref{Lead+C}). 
}
\item{
Proceeding in the same way for $v\rightarrow 0$ we get
\be\label{Asy+3}
{\cal F}(n,k,\frac{n+1}{2}|w,v)=(-1)^{-n/2}\frac{k}{n}v^{-\frac{n+1}{2}+n-k}
w^{\frac{1}{2}-n+k}+O(v^{\frac{n-1}{2}-k+1})
\ee
}
\item{
Knowing the poles of the analytic function ${\cal F}$ as well as the residues
(\ref{Asy+1}), (\ref{Asy+2}), (\ref{Asy+3}), we can determine the
complete function
\be
{\cal F}(n,k,\frac{n+1}{2}|v,w)=(v-w)^{-n/2}w^{-\frac{n+1}{2}+k}
v^{-\frac{n+1}{2}+n-k}\left(\frac{n-k}{n}v+\frac{k}{n}w\right).
\ee
}
\end{enumerate}
To compare with (\ref{ADefFuCrRSpec}) we substitute the value of 
${\cal F}$
in  (\ref{++ratio}) :
\bea
&&\frac{\langle S_{n-k,k}(v)S^\dagger_{n-k,k}(w)
\sigma^\frac{n+1}{2}_n(0)\sigma^{\frac{n+1}{2}\dagger}_n(\infty)\rangle}
{\langle\sigma^\frac{n+1}{2}_n(0)\sigma^{\frac{n+1}{2}\dagger}_n(\infty)
\rangle}=
w^{-n/2}x^{-n/12-(n+1)/6}(x-1)^{1/3}\nonumber\\
&&\qquad\times (x-1)^{-n/2-1/3}x^{3n/4-1/3-k}\frac{(n-k)x+k}{n}
\eea
Thus we finally find the function $f$ which depends on the cross ratio (see (\ref{DefFIndFin})):
\be\label{FFact++}
f(n,k,k|\frac{n+1}{2},0;\frac{n+1}{2},0|x)=
(-1)^{n+1}(x-1)^{-n/2-1/3}x^{3n/4-1/3-k}\frac{(n-k)x+k}{n}.
\ee

\subsection{Evaluation of the basic correlator with $P=Q=\frac{n-1}{2}$.}

%Let us briefly mention the remaining case of the correlator
%(\ref{BasicCorr1}). 
As in
the previous subsection we define:
\be
{\cal F}(n,k,\frac{n-1}{2}|v,w)\equiv
\frac{\langle S_{n-k,k}(v)S^\dagger_{n-k,k}(w)
\sigma^\frac{n-1}{2}_n(0)\sigma^{\frac{n-1}{2}\dagger}_n(\infty)\rangle}
{\langle\sigma^\frac{n-1}{2}_n(0)\sigma^{\frac{n-1}{2}\dagger}_n(\infty)
\rangle}
\ee
The asymptotics of this expression are:
\bea
{\cal F}(n,k,\frac{n-1}{2}|v,w)&=&\frac{1}{(v-w)^{n/2}}+
O\left(\frac{1}{(v-w)^{n/2-1}}\right),\\
{\cal F}(n,k,\frac{n-1}{2}|v,w)&=&(-1)^{-n/2}\frac{n-k}{n}
v^{-\frac{n+1}{2}+n-k}
w^{\frac{1}{2}-n+k}+O(v^{\frac{n-1}{2}-k+1}),\\
{\cal F}(n,k,\frac{n-1}{2}|w,v)&=&\frac{k}{n}w^{-\frac{n+1}{2}+k}
v^{\frac{1}{2}-k}+O(w^{-\frac{n+1}{2}+k+1})
\eea
This produces the unique expression for functions ${\cal F}$ and $f$: 
\bea
{\cal F}(n,k,\frac{n-1}{2}|w,v)&=&(v-w)^{-n/2}w^{-\frac{n+1}{2}+k}
v^{-\frac{n+1}{2}+n-k}\left(\frac{k}{n}v+\frac{n-k}{n}w\right),\nonumber\\
\label{FFact--}
f(n,k,k|\frac{n-1}{2},0;\frac{n-1}{2},0|x)&=&
(-1)^{n-1}(x-1)^{-n/2+1/3}x^{3n/4-2/3-k}\frac{kx+n-k}{n}
\eea

\subsection{Evaluation of the basic correlator for $P\ne Q$.}

To evaluate the correlation function (\ref{BasicCorr2}), we define a function of two complex variables:
\bea
{\cal G}(n,k,\frac{n-1}{2}|v,w)\equiv
\frac{\langle S_{n-k,k}(v)S^\dagger_{n-k-1,k+1}(w)\sigma^{\frac{n-1}{2}}_n(0)
\sigma^{{\frac{n+1}{2}}\dagger}_n(\infty)\rangle}
{\langle\sigma^{\frac{n+1}{2}}_n(0)
\sigma^{{\frac{n+1}{2}}\dagger}_n(\infty)\rangle}.\nonumber
\eea
In the limit of $w\rightarrow 0$ one has an expansion
\bea
{\cal G}(n,k,\frac{n-1}{2}|v,w)=
\sum_i C_i(n,k+1,\frac{n-1}{2}){\bar C_i(n,k,\frac{n+1}{2})}
w^{-\frac{n-1}{2}-\frac{n}{4}+\Delta_i}
v^{\frac{n-1}{2}-\frac{n}{4}+1-\Delta_i},\nonumber
\eea
and the leading contribution to the right--hand side comes from the exchange by 
${\cal A}_{n,k}$, which is the lightest operator produced in the OPE $S^\dagger_{n-1-k,k+1}(w)\sigma^{\frac{n-1}{2}}_n(0)$.  
The fusion rules (\ref{SpinTwistOPE}) give the leading singularity in ${\cal G}$,
\be
{\cal G}(n,k,\frac{n-1}{2}|v,w)=-\frac{\sqrt{(k+1)(n-k)}}{n}
w^{-(n-1)/2+k}v^{1/2-k}+O(w^{-n/2+3/2+k}).
\ee
Applying similar arguments in the limit of small $v$, we get:
\be
{\cal G}(n,k,\frac{n-1}{2}|w,v)=\frac{\sqrt{(k+1)(n-k)}}{n}
(-1)^{-n/2}v^{n/2-1/2-k}w^{3/2-n+k}+O(v^{n/2+1/2-k}).
\ee
To find the order of the pole\footnote{It turns out that we do not need its residue.}
at $v=w$, we look at the OPE
\bea
&&S_{n-k,k}(v)S^\dagger_{n-k-1,k+1}(w)\sim
\left[\left(J^-_0\right)^k S_{n,0}(v)\right]
\left[\left(J^+_0\right)^{k+1} S^\dagger_{n,0}(v)\right]\nonumber\\
&&\qquad\sim \left(J^-_0\right)^k
\left\{S_{n,0}(v)\left[J^+_0 S^\dagger_{n,0}(w)\right]\right\}
\sim (v-w)^{-n/2+1}.
\eea
In this expression we dropped all numerical coefficients and at the last step
we also used the expression for $J^+_0 S^\dagger_{n,0}(w)$,
\bea
J^+_0 S^\dagger_{n,0}(w)&=&J^+_0
\prod_{j=1}^n:\exp\left(\frac{i}{2}e_a\phi_j^a(z)\right):\nonumber\\
&=&
:\exp\left(-\frac{i}{2}e_a\phi_1^a(z)\right)
\prod_{j=2}^n\exp\left(\frac{i}{2}e_a\phi_j^a(z)\right):~+~
\mbox{permutations}\nonumber
\eea
Collecting information about the limits of ${\cal G}$, we finally get
\bea\label{HetResult}
&&\frac{\langle S_{n-k,k}(v)S^\dagger_{n-k-1,k+1}(w)\sigma^{\frac{n-1}{2}}_n(0)
\sigma^{{\frac{n+1}{2}}\dagger}_n(\infty)\rangle}
{\langle\sigma^{\frac{n+1}{2}}_n(0)
\sigma^{{\frac{n+1}{2}}\dagger}_n(\infty)\rangle}\\
&&\qquad\qquad=
-\frac{\sqrt{(k+1)(n-k)}}{n}w^{-n/2+1}\left(\frac{v}{v-w}\right)^{n/2-1}
\left(\frac{v}{w}\right)^{1/2-k}\nonumber\\
&&\qquad\qquad\equiv(-1)^{-n}w^{1-n/2}x^{-n/4+1/2}f(n,k,k+1|\frac{n-1}{2},0;\frac{n+1}{2},0|x).
\nonumber
\eea
The last line is an application the general relation (\ref{ADefFuCrRSpec}) to the present case. 
This leads to the final expression for function $f$:
\be\label{FFunc+-}
f(n,k,k+1|\frac{n-1}{2},0;\frac{n+1}{2},0|x)=(-1)^{n+1}
\frac{\sqrt{(k+1)(n-k)}}{n}x^{3n/4-1-k}(x-1)^{-n/2+1}.
\ee

\section{The results for the four point functions.}
\label{SectResult}
\renewcommand{\theequation}{7.\arabic{equation}}
\setcounter{equation}{0}

In the previous section we have evaluated the basic correlators containing
two spin and two twist operators, i.e. we considered the special case of
(anti)chiral twist operators $\sigma_n^Q$ and $\sigma_n^{Q\dagger}$. To obtain
the result for the general four point functions containing
two spin and two twist operators one can take the result of the previous
section and use some combinatorics. The details are presented in the
appendices \ref{SectChirTwist} and \ref{SectGenCorr}.

We have been focusing on the case when the order of
the twist operator $\sigma_n$ is the same as the order of permutation group
$S_n$. As discussed in section \ref{SectGen4ptRed}, in the more general case of 
$S_N$ orbifold with $N>n$, the answer can be
written as a series of correlation functions on $S_n$ orbifold. However in
the case of the special correlator (\ref{DeflEq0})
the series (\ref{TempLargeN}) contains only one term,
and in this section we will present the four point
functions for this case.
\begin{enumerate}[1.]
\item{{\bf The result for $P=Q=\frac{n+1}{2}$.}\\
Performing the combinatorial analysis presented in the appendix \ref{AppE2}, one deduces the expression (\ref{FinRes++2}) for
the relevant function $f$, which depends on the cross--ratio (\ref{DefCrRat}). Using the general relation (\ref{DefFIndFin}) between such function and the corresponding correlator on $S_n$ orbifold and the reduction formula (\ref{TempLargeN}), we find}
\end{enumerate}
\bea\label{LargeN++}
&&\langle S_{N,0}(v)S^\dagger_{N-k,k}(w)\left((J_0^-)^p
\sigma^{\frac{n+1}{2}}_n\right)(0)
\left((J_0^+)^q\sigma^{{\frac{n+1}{2}}\dagger}_n\right)(a)\rangle
=(-1)^{q+n+k+1}\nonumber\\
&&\qquad\times\left(\frac{(N-k)!k!}{N!}\right)^\frac{1}{2}
\frac{(k+q)!(n-1)!}{k!(n+1-k-q)!}(v-w)^{-N/2}
\left(\frac{vwa^2(a-v)(a-w)}{(v-w)^2}\right)^{-n/4-1/6}
\nonumber\\
&&\qquad\times 
(x-1)^{-n/2-1/3}
x^{3n/4-1/3}
\left(\frac{1+x}{x}\right)^{k}
\\
&&\qquad\times\left.
%\left(-\frac{1+x}{x}\right)^{k}
\left(n+1-k-y\frac{\partial}{\partial y}\right)
\left\{nx+k\frac{1-x}{1+x}+
\frac{1-x}{1+x}y\frac{\partial}{\partial y}\right\}
(1+y)^q\right|_{y=-\frac{1+x}{x}}\nonumber
\eea
\begin{enumerate}[2.]
\item{{\bf The result for $P=Q=\frac{n-1}{2}$.}\\
Proceeding as before and using (\ref{FinRes--1}) instead of
(\ref{FinRes++2}), we get}
\end{enumerate}
\bea\label{LargeN--}
&&\langle S_{N,0}(v)S^\dagger_{N-k,k}(w)\left((J_0^-)^p
\sigma^\frac{n-1}{2}_n\right)(0)
\left((J_0^+)^q\sigma^{\frac{n-1}{2}\dagger}_n\right)(a)\rangle
=(-1)^{q+n+k+1}\nonumber\\
&&\qquad\times\left(\frac{(N-k)!}{N!}\right)^\frac{1}{2}
\frac{(k+q)!}{\sqrt{k!}}
\frac{(n-1)!(n-k)^{-1}}{(n-1-k-q)!}\nonumber\\
&&\qquad\times
(v-w)^{-N/2}%\nonumber\\
%&&\qquad\times
\left(\frac{vwa^2(a-v)(a-w)}{(v-w)^2}\right)^{-n/4+1/6}\nonumber\\
&&\qquad\times (x-1)^{1/3-n/2}x^{3n/4-2/3}
\left(\frac{1+x}{x}\right)^{k}\\
&&\qquad\times
\left.\left\{n+\frac{x-1}{x+1}\left(k+y
\frac{\partial}{\partial y}\right)\right\}
F(-q,k{-}l{-}n;k{-}l{-}n{+}1;y)\right|_{y=-\frac{1+x}{x}}\nonumber
\eea
\begin{enumerate}[3.]
\item{{\bf The result for $P=\frac{n-1}{2}$, $Q=\frac{n+1}{2}$.}\\
Equation (\ref{FinRes+-}) gives}
\end{enumerate}
\bea\label{LargeN+-}
&&\langle S_{N,0}(v)S^\dagger_{N-k,k}(w)\left((J_0^-)^p
\sigma^\frac{n-1}{2}_n\right)(0)
\left((J_0^+)^q\sigma^{\frac{n+1}{2}\dagger}_n\right)(a)\rangle
=(-1)^{q+k+n}\nonumber\\
&&\qquad\times\left(\frac{(N-k)!}{N!}\right)^\frac{1}{2}
\frac{p!(n-1)!}{(n-1-p)!\sqrt{k!}}\nonumber\\
&&\qquad\times(v-w)^{-N/2}
\left(\frac{vwa^2(a-v)(a-w)}{(v-w)^2}\right)^{-n/4}
\left(\frac{vw}{(a-v)(a-w)}\right)^{1/2}\\
&&\qquad\times 
(x-1)^{-n/2+1}x^{3n/4-1}\phantom{\frac{e^x}{f^y}}
%&&\qquad\times
\left.\left(\frac{1+x}{x}\right)^{k-1}
\left(k+y\frac{\d}{\d y}\right)(1+y)^q\right|_{y=-\frac{1+x}{x}}
\nonumber
\eea

%==========================================================================
%
%==========================================================================

\section{The physical amplitude}
\label{SectGrav}
\renewcommand{\theequation}{8.\arabic{equation}}
\setcounter{equation}{0}

After evaluating the CFT amplitude that corresponds to the absorption and subsequent emission of a quantum, we would like to place this computation in its physical context. The initial state $|i\rangle$ depicted in figure \ref{figGround}(a) represents a D1--D5 bound state obtained by wrapping D1 branes on $S^1$ and D5 branes on $S^1\times T^4$. This bound state gives an effective string in the direction $S^1$, and we assume that the radius $R_y$ of this $S^1$ is large. The low energy dynamics of the bound state gives  a 1+1 dimensional CFT. Figure \ref{figGround}(b) represents the correlator in this theory that is relevant to our process of interest. Section \ref{SectResult} presents the results for the required 4-point function in the plane. In section \ref{SectCylnd} we will carry out  the map to the cylinder which is depicted in figure \ref{figGround}(b). 

The correlation function on the cylinder is related to the gravity amplitude depicted in figure \ref{figGround}(a).
The full spacetime evolution takes place in a six--dimensional space containing time $t$, the $y$ coordinate along $S^1$, the radial direction $r$, and the angular coordinates 
$(\theta,\psi,\phi)$ that make an $S^3$ surrounding the effective string\footnote{We have dimensionally reduced on the $T^4$ since nothing depends on the torus directions.}. For convenience, we place a boundary at some large radius $r=R_1$, and the limit 
$R_1\r\infty$ will be taken at the end. The geometry created by the branes is flat space for $R_1>r\gg \sqrt{Q}$, there is a `neck' region for $r\sim \sqrt{Q}$, and an $AdS_3\times S^3$ region for $r\ll \sqrt{Q}$. 

We wish to consider a process where quanta are incident onto the effective string from the flat spacetime region of large $r$. Such a quantum can get absorbed by the effective string, and after some time there can be a re-emission process where a quantum is radiated back to the flat spacetime region. The physics of the effective string is captured by the CFT. Our task is to relate the amplitude computed in the CFT (on the cylinder) to the full physical process of interest.  In section \ref{SectCollFact} we will use the general procedure developed in \cite{acm} which relates CFT correlators to the full process of emission/absorption from the D-brane system.

%==========================================================================
%
%==========================================================================
\subsection{Map to the cylinder.}
\label{SectCylnd}

We begin by translating the amplitude on the plane given in section \ref{SectResult} to the CFT amplitudes on the cylinder shown in figure \ref{figGround}(b). In the complex plane, the Ramond vacua are described by the insertion of appropriate spin operators at definite points. In the cylinder description, the Ramond vacuum states $|i\rangle$ and $|f\rangle$ are represented by specifying appropriate boundary conditions on the two ends of the cylinder depicted in figure \ref{figGround}(b). 

We take the  map from the cylinder coordinate $w$ to the sphere coordinate $z$,
\bea
z=e^w.
\eea
Under this map an operator $B$ on the plane goes to an operator on the cylinder; we denote the latter by  $\tilde B$. 
Starting with a general four point function on the sphere,
\be\label{cyl1}
\langle S_{N,0}(z_1)S^\dagger_{N-k,k}(z_2)B_3(z_3)B_4(z_4)\rangle,
\ee
we consider the limit when the spin operators in (\ref{cyl1})
are placed at $w=\pm\infty$ on the cylinder. Then we have
\be\label{CylCoordLim}
w_1\rightarrow -\infty,\qquad  w_2\rightarrow\infty,\qquad x\rightarrow \frac{z_4}{z_3}=e^{w_4-w_3}\,.
\ee
Using the standard transformation rules for the operators in CFT, we find the expression for a 
two--point function on the cylinder with specific boundary conditions at $w\r\pm\infty$:
$$
\langle N,0|{\tilde B}_3(w_3){\tilde B}_4(w_4)|N-k,k\rangle\equiv
\lim_{w_{2,1}\rightarrow \pm\infty}
\langle {\tilde S}_{N,0}(w_1){\tilde S}^\dagger_{N-k,k}(w_2)
{\tilde B}_3(w_3){\tilde B}_4(w_4)\rangle e^{-\Delta_1w_1+\Delta_2w_2}
$$
\bea\label{Cyl2pt}
=(-1)^{\Delta/3-2\Delta_1-\Delta_3-\Delta_4}
e^{(\Delta/3-\Delta_1)(w_3+w_4)}
\left(e^{w_4}-e^{w_3}\right)^{\Delta/3-\Delta_3-\Delta_4}f_N(e^{w_4-w_3}).
\eea
Here $\Delta_1=\Delta_2=N/4$, $\Delta=N/2+\Delta_3+\Delta_4$, $\Delta_3$ and 
$\Delta_4$ are dimensions of $B_3$ and $B_4$. We are interested in the case where $B_3$ and $B_4$ are rotated twist operators of order $n$. Then equation  (\ref{TempLargeN}) reduces the calculation to one for the case where we have an
$S_n$ orbifold
\bea\label{fLrgN}
f_N(e^{w_4-w_3})=\left[\frac{(N-k)!n!}{N!(n-k)!}\right]^{\frac{1}{2}}f(e^{w_4-w_3})
\eea
and the functions $f$ are given by (\ref{FinRes++2}) (\ref{FinRes--1}), (\ref{FinRes+-}). In the following subsections, we will express the gravity amplitude in terms of the two--point function (\ref{Cyl2pt}). 

\subsection{Evaluation of the gravity amplitude}
\label{SectCollFact}

In this subsection we will use the general analysis of \cite{acm} to extract the gravitational amplitude shown in figure \ref{figGround}(a). For details of notation etc. we refer the reader to \cite{acm}.

The supergravity quanta are absorbed and emitted in definite partial waves, so we begin by 
expanding the relevant supergravity field in spherical modes:
\bea
\hat\phi&=&\sum_{p,k} \sum_{l, m, m'}\Big ({1\over \sqrt{2\omega_{p,k}}}\phi_{p,k}^{lmm'}(r)Y_{l m m'}(\theta,\psi,\phi)e^{-i\omega_{p,k} t} \hat a_{p,k} \nn
&+& {1\over \sqrt{2\omega_{p,k}}}(\phi_{p,k}^{lmm'})^*(r)Y^*_{l m m'}(\theta,\psi,\phi)e^{i\omega_{p,k} t} \hat a_{p,k}^\dagger\Big  )
\eea
Here $Y_{lmm'}$ are the tensor spherical harmonics for the appropriate supergravity field, which are normalized by
\be
\int |Y_{lmm'}|^2d\Omega=1,
\ee
and indices $n, k$ correspond to the radial quantum number and the momentum in the $y$ direction respectively. 

To make contact with the CFT description, place a boundary surface at a location $r=r_b$ inside the $AdS_3$ region (i.e., $r_b\ll \sqrt{Q}$). The coupling of the gravity quantum to the corresponding operator $\hat O$ in the CFT is then \cite{acm}
\be
S_{int}=-c_l\int \sqrt{-g_2}dtdy [\p^l\phi_{p,k}^{lmm'}]\Big |_{r=0}\hat O(t,y)
\label{interact}
\ee
where $g={r_b^2\over Q}$ is the determinant of the 2-d metric on the CFT surface, 
and $[\p^l\phi_{p,k}^{lmm'}]\Big |_{r=0}$ is defined so that as $r\r 0$
\be
\phi_{p,k}^{lmm'}\approx [\p^l\phi_{p,k}^{lmm'}]\Big |_{r=0} r^l Y_{lmm'}(\Omega)
\ee
The constant $c_l$ is
\be
c_l=\left[\frac{Q^{2h}(2h-1)^2(2\pi)^4 V}{8\pi^2 G }\right]^\frac{1}{2} r_b^{2h-2}
\label{cl}
\ee
where we note that for the gravity modes under consideration, the field is undistorted in the `neck' and so the factor $b_l$ in \cite{acm} is unity.

Consider a supergravity quantum in the gravity mode $\phi_{p,k}^{lmm'}$.  This quantum can be absorbed into the CFT degrees of freedom through the operator 
$\hat O$. At some later time, another gravity mode 
$\phi_{\t p,\t k}^{l\t m\t m'}$ can be emitted from the CFT in the interval $dy_2dt_2$ around the point $(y_2,t_2)$. Let us denote the amplitude for such two--step process by ${\cal A}$.
To evaluate this amplitude, we need the following ingredients.
\begin{enumerate}
\item{The interaction (\ref{interact}) gives a 2-point function $\langle O(y_1,t_1)\t O(y_2,t_2)\rangle$, which has been evaluated in the last subsection\footnote{One should also specify the appropriate initial and final Ramond ground states as in (\ref{Cyl2pt}), but we omit this specification to shorten the expressions.}. However, we should note that (\ref{Cyl2pt}) gives the correlator on a unit cylinder with circumference $2\pi$, while here we work with CFT on the $y$ circle with length $L_y=2\pi {r_b\over \sqrt{Q}} R_y$ (see \cite{acm}). This gives
\bea
\langle O(y_1,t_1)\t O(y_2,t_2)\rangle=
{1\over [{r_b\over \sqrt{Q}}|y_1-y_2|]^{4h}}
\langle O(y_1,t_1)\t O(y_2,t_2)\rangle_{unit}
\eea
where the subscript `unit' on the rhs refers to the correlator (\ref{Cyl2pt}) on the unit cylinder.
}
\item{The absorption and emission of quanta happen due to 
the coupling (\ref{interact}); this gives the factor 
\be
[\p^l\phi_{p,k}^{lmm'}]\Big |_{r=0} [\p^l\phi_{\t p,\t k}^{l\t m\t m'}]\Big |_{r=0} c_l^2\left ( {r_b^2\over Q}\right )^2 
\ee
in the gravity amplitude. Here $c_l$ given by (\ref{cl}).
}
\item{The field operators for the gravity field give the phase space factors
\be
{1\over\sqrt{2\omega_{p,k}}}{1\over \sqrt{2\omega_{\t p,\t k}}}e^{-i\omega_{p,k} t_1} e^{i k y_1}e^{i\omega_{\t p,\t k}t_2} e^{-i\t k y_2}dy_1dt_1dy_2dt_2
\ee
}
\end{enumerate}

\noindent
Putting all this together, we find
\bea\label{GravAmpl}
{\cal A}&=&{1\over\sqrt{2\omega_{p,k}}}{1\over \sqrt{2\omega_{\t p,\t k}}}e^{-i\omega_{p,k} t_1} e^{i k y_1}e^{i\omega_{\t p,\t k}t_2} e^{-i\t k y_2}dy_1dt_1dy_2dt_2\nn
&\times& {1\over |y_1-y_2|^{4h}} Q^{4h-2}{(2h-1)^2(2\pi)^4 V\over 8\pi^2 G } [\p^l\phi_{p,k}^{lmm'}]\Big |_{r=0} [\p^l\phi_{\t p,\t k}^{l\t m\t m'}]\Big |_{r=0}\nn
&\times& 
\langle O(y_1,t_1)\t O(y_2,t_2)\rangle_{unit}
\eea
Here $h$ is the dimension of the twist, and the 2-point function between the initial and final Ramond ground states is given by (\ref{Cyl2pt}). We recall that equations (\ref{Cyl2pt}) and (\ref{fLrgN}) refer to our main results (\ref{FinRes++2}) (\ref{FinRes--1}), (\ref{FinRes+-}) (or, alternatively, (\ref{LargeN++}), (\ref{LargeN--})).

To get the final correlator, we put the supergravity quanta in wavepackets around the locations $y_1, y_2$; these wavepakets are narrow in the $y$ direction (where their width is much smaller than $|y_1-y_2|$) and in the the momentum space (where they are picked around $k=0, \t k=0$ with width $\Delta$)\footnote{These requirements are compatible since we take $|y_1-y_2|$ large compared to the wavelength of the supergravity quanta.}.  This leads to the physical amplitude
\be
{\Delta^2\over \pi}\int_{-\infty}^\infty \int_{-\infty}^\infty dk d\t k \exp\left[-{k^2\over \Delta^2}-{\t k^2\over \Delta^2}\right]{\cal A}
\ee
To recap, this expression gives  the amplitude for the process where a supergravity quantum in one spherical harmonic  hits the D1--D5 bound state in a wavepacket localized around  $y=y_1$, and emerges with a different harmonic in  a wavepacket localized around  $y=y_2$. 

\subsection{Large $N$ behavior of the amplitude}
\label{SubsLrgN}
Although the two--point function entering the gravity amplitude 
(\ref{GravAmpl}) is rather complicated, an important physical effect can be extracted from its scaling with $N$, which is given by the combinatorial factor in (\ref{fLrgN}):
\bea\label{LrgNmain}
{\cal A}_{k,{\bar k}}\sim  N^{-(k+{\bar k})/2}
\eea
In particular, we find that the amplitude for going to the final state rotated by $k={\bar k}$ units is suppressed as $N^{-k}$. 
In this paper we considered a transition to a very special state (which is a rotation of the initial vacuum), but if such suppression persists for more general final states, it would have important consequences for the dynamics of black holes: it would imply that are given state of a hole cannot transition to all $e^S$ states. 

To investigate this question further, in appendix \ref{AppLrgN} we analyzed a large $N$ behavior of a more general amplitude. As in (\ref{Cyl2pt}), we took the initial Ramond vacuum to be the highest weight state with $j={\bar j}=\frac{N}{2}$, but we allowed the final vacuum to have different $SU(2)\times SU(2)$ quantum numbers\footnote{The initial state can {\it always} be chosen to be the highest weight state by performing SU(2)$\times$SU(2) rotation, so the only assumption is $j={\bar j}=\frac{N}{2}$.}. Notice that the final state in (\ref{Cyl2pt}), (\ref{fLrgN}) has
\bea
j=\frac{N}{2},\quad j_3=\frac{N}{2}-k;\qquad
{\bar j}=\frac{N}{2},\quad {\bar j}_3=\frac{N}{2}-{\bar k}.
\eea
In appendix \ref{AppLrgN} we analyzed the transitions to the Ramond vacua with
\bea
j,\quad j_3=\frac{N}{2}-k;\qquad
{\bar j},\quad {\bar j}_3=\frac{N}{2}-{\bar k}.
\eea
and we found that only states with $j={\bar j}$ are allowed, and the scaling of the transition amplitude is given by (\ref{FinConj}):
\bea\label{FinConjMain}
{\cal A}_{j,k;j,{\bar k}}\sim N^{-(k+{\bar k})/2}
N^{(\frac{N}{2}-j)/2}
\eea
Since $j_3\le j$ and ${\bar j}_3\le {\bar j}_3$, we find that 
\bea
j\ge \frac{N}{2}-\mbox{min}(k,{\bar k}),
\eea 
and for given $k,{\bar k}$ the minimal suppression happens when the last inequality is saturated. For example, if 
$k={\bar k}$, then the minimal suppression is
\bea\label{MinSuppr}
{\cal A}_{\frac{N}{2}-k,k;\frac{N}{2}-k,k}\sim N^{-k/2}.
\eea

To summarize, we found that a transition from a Ramond vacuum with $j_0={\bar j}_0=\frac{N}{2}$ can only happen to a vacuum with $j={\bar j}$, and the relevant amplitude scales as (\ref{FinConjMain}), where $k=|\Delta j_3|$, 
${\bar k}=|\Delta {\bar j}_3|$. For a given value of 
$k={\bar k}$, the suppression becomes minimal if $j=\frac{N}{2}-k$, and it is given by (\ref{MinSuppr}).

\section{Discussion}

One of the most important features of a black hole is a large number of degenerate states. Thus a quantum can fall into a hole in one state and emerge later in some other mode, while changing the state of the black hole as well. To study the dynamics of the Hawking radiation, one should get a better understanding of this absorption/emission process, in particular, one should determine whether the full space of degenerate states is being explored, or gravitational quanta can only cause transitions between `nearby states' of the black hole.  

The simplest absorption/emission process does not change states of the black hole and of the infalling quantum. The relevant amplitude can be computed by solving the wave equation for the quantum on the background geometry produced by the black hole microstate; for the state discussed here, this computation was done in \cite{hottube}. For more complicated processes, where the quantum and the black hole state change their spins, the gravity computation becomes more involved since it requires a gravitational coupling vertex. Rather than solving this problem at strong coupling, we analyzed its counterpart at weak coupling by going to the dual CFT, and one may hope that general features of the scattering matrix would be similar in the two limits. It would be very interesting to
perform an independent gravity calculation and to see whether a remarkable agreement between the CFT and gravity correlators \cite{lmtwo,GKDP} persists for the amplitudes derived here. Alternatively, one can try to find corrections to our results caused by the deformation from the free orbifold CFT following the ideas of \cite{gava}.

In this paper we studied transitions between the Ramond vacuum with largest allowed spin and its SU(2)$\times$SU(2) rotations. We found that the black hole does not jump between different degenerate states with equal probability. In particular, the amplitude for the spin of the hole to change by $k$ units is suppressed by $1/N^{k}$. In section \ref{SubsLrgN} we studied the large $N$ scaling for more general transitions from the maximally rotated state, and we found that for the given value of $k$, it is the transition to the state with the lowest allowed angular momentum that dominates, although its amplitude is still suppressed by $1/N^{k/2}$ relative to the trivial case. These results are in line with the expectation that the transition which does not alter the state of the hole has the maximal amplitude: this case corresponds to propagation of the infalling quantum on the background geometry of the microstate. Scaling of amplitudes by different powers of $N$ suggests that some pairs of microstates are `close', and other pairs are `far away' in the set of all Ramond vacua, and it would be interesting to find an `effective metric' to quantify this distinction by extending our results to arbitrary initial states. 

We should note that the initial state studied in this paper is very special: it has the maximal spin allowed for the D1--D5 system, so any transition to another state is accompanied by a change in angular momentum, which eventually led to the suppression of the corresponding amplitude. In the generic case, 
we expect to find more Ramond vacua in the `vicinity' of the initial state, so it should become very {\it easy} for the black hole to change its state when hit by an infalling quantum. It would be interesting to repeat our analysis for generic initial states and to find the `effective metric' on the space of all Ramond vacua.

\section*{Acknowledgements}

The work of SDM was supported in part by DOE grant DE-FG02-91ER-40690.

\appendix

\section{Reduction from the $S_N$ orbifold theory to an $S_n$
orbifold theory.}
\label{AppRdct}
\renewcommand{\theequation}{A.\arabic{equation}}
\setcounter{equation}{0}

In this appendix we will derive the relation (\ref{TempLargeN}) which reduces the evaluation of correlation function (\ref{def4pointB}) on $M^N/S_N$ orbifold to computation in $M^n/S_n$ theory.

In the correlation function (\ref{genSSss1}) the spin
fields
$S$ change the boundary condition for fermionic variables from NS to
R for all $N$ copies
of the theory. The twist operators, on the other hand,  permute only
$n$ copies of the $c=6$ CFT. It would be desirable to derive the expression for
(\ref{genSSss1}) which contains only the
$n$ copies involved in the permutation. Let us assume that
the twist operators $\sigma_n$  permute the fields
$\phi^a_1(z),\dots,\phi^a_n(z)$. Then
it is convenient to split the operator $S_{N,0}(z)$ involved in the definition
(\ref{defGenerSpin}) of the $S_{N-k,k}$ into two separate parts:
\be
S_{N,0}(z)=\prod_{j=1}^n:\exp\left(\frac{i}{2}e_a\phi_j^a(z)\right):~
\prod_{j=n+1}^N:\exp\left(\frac{i}{2}e_a\phi_j^a(z)\right):
\ee
where we have used the definition (\ref{defSpecSpin}) of $S_{N,0}$.
Let us introduce the
following  shorthand notation for the two terms involved in the above expression:
\bea\label{SpinUpPer}
{\tilde S}^{(1)}_{n,0}(z)&\equiv&
\prod_{j=1}^n:\exp\left(\frac{i}{2}e_a\phi_j^a(z)\right):,\\
\label{SpinUpNPer}
{\tilde S}^{(2)}_{N-n,0}(z)&\equiv&
\prod_{j=n+1}^N:\exp\left(\frac{i}{2}e_a\phi_j^a(z)\right):
\eea
These two operators have a simple meaning. Operator (\ref{SpinUpPer}) is a
spin operator in CFT formed by the fields $\phi^a_1(z),\dots,\phi^a_n(z)$,
and (\ref{SpinUpNPer}) is a spin operator in CFT containing the remaining
copies: $\phi^a_{n+1}(z),\dots,\phi^a_N(z)$. These are the special spin
operators which have a maximal value of $j_3$: $j_3=j=n/2$ for
${\tilde S}^{(1)}_{n,0}$ and $j_3=j=(N-n)/2$ for ${\tilde S}^{(2)}_{N-n,0}$.
As in (\ref{defGenerSpin}) we can construct more general spin operators for
both
conformal field theories. To do this we first split the diagonal components of
the currents (\ref{generatorS}) into two contributions: one coming from the
first $n$
copies and one coming from the remaining $N-n$ copies of the theory. In
particular, the expression for $J^-_0$ becomes:
\be
J^-(z)=\sum_{j=1}^n\exp\left(-ie_a\phi^a_j(z)\right)+
\sum_{j=n+1}^N\exp\left(-ie_a\phi^a_j(z)\right)\equiv
J^{(1)-}(z)+J^{(2)-}(z)
\ee
Then we can define the spin operators for the two CFTs in the same manner as
we did it for the complete theory in (\ref{defGenerSpin}):
\bea
{\tilde S}^{(1)}_{n-k,k}(z)&=&
\sqrt{\frac{(n-k)!}{k!n!}}\left(J^{(1)-}_0\right)^k
{\tilde S}^{(1)}_{n,0}(z),\\
{\tilde S}^{(2)}_{N-n-k,k}(z)&=&
\sqrt{\frac{(N-n-k)!}{k!(N-n)!}}\left(J^{(2)-}_0\right)^k
{\tilde S}^{(1)}_{N-n,0}(z).
\eea
We can express $S_{N-k,k}$ given by  (\ref{defGenerSpin}) in
terms of the above operators:
\bea\label{largeNSum}
&&S_{N-k,k}(z)=\sqrt{\frac{(N-k)!}{k!N!}}\left(J^-_0\right)^k
\left({\tilde S}^{(1)}_{n,0}(z) {\tilde S}^{(2)}_{N-n,0}(z)\right)\\
%&&\qquad=
%\sqrt{\frac{(N-k)!}{k!N!}}\sum_{b=0}^k\frac{k!}{b!(k-b)!}
%\left\{\left(J^-_0\right)^{k-b}
%{\tilde S}^{(1)}_{n,0}(z)\right\}
%\left\{\left(J^-_0\right)^{b}{\tilde S}^{(2)}_{N-n,0}(z)\right\}\\
&&=\sqrt{\frac{(N-k)!k!}{N!}}\sum_{b=0}^k
\left(\frac{n!(N-n)!}{(n+b-k)!(N-n-b)!b!(k-b)!}\right)^{1/2}
{\tilde S}^{(1)}_{n+b-k,k-b}(z){\tilde S}^{(2)}_{N-n-b,b}(z)\nonumber
\eea

Let us now use the expansion (\ref{largeNSum}) to simplify the correlator
(\ref{genSSss1}). Note that the two sets of fields:
$(\phi_1^a,\dots,\phi_n^a)$ and
$(\phi_{n+1}^a,\dots,\phi_N^a)$ decouple.  We get
\bea\label{Expand4ptN}
&&\langle S_{N-l,l}(v)
S_{N-k,k}^\dagger(w)\left((J_0^-)^p\sigma^P_n\right)(0)
\left((J_0^+)^q\sigma^{Q\dagger}_n\right)(a)\rangle_{S_N}
=\left(\frac{(N{-}k)!k!}{N!}\frac{(N{-}l)!l!}{N!}\right)^{1/2}\nonumber\\
&&\qquad\times\sum_{b=0}^{k}\sum_{c=0}^{l}
\left(\frac{n!(N-n)!}{(n{+}b{-}k)!(N{-}n{-}b)!(k{-}b)!b!}\right)^{1/2}
\left(\frac{n!(N-n)!}{(n{+}c{-}l)!(N{-}n{-}c)!(l{-}c)!c!}\right)^{1/2}
\nonumber\\
&&\qquad\qquad\times
\langle {\tilde S}^{(1)}_{n+c-l,l-c}(v)
{\tilde S}^{(1)\dagger}_{n+b-k,k-b}(w)
\left[(J_0^-)^p\sigma^P_n\right](w)
\left[(J_0^+)^q\sigma^{Q\dagger}_n\right](a)\rangle_{S_n}\nonumber\\
&&\qquad\qquad \times\langle 
{\tilde S}^{(2)}_{N-n-c,c}(v)
{\tilde S}^{(2)\dagger}_{N-n-b,b}(w)\rangle_{S_{N-n}}
\eea
We have now obtained a sum  of terms, where  each term is a  product of a
correlator from the first $n$ copies of the $c=6$ CFT and a
correlator from the remaining
$N-n$ copies.

Using  charge conservation and the normalization
(\ref{SpinNormal}), we note:
\be
\langle 
{\tilde S}^{(2)}_{N-n-c,c}(v)
{\tilde S}^{(2)\dagger}_{N-n-b,b}(w)\rangle_{S_{N-n}}=
\delta_{b,c}(v-w)^{-(N-n)/2}
\ee
Thus we see that the evaluation of our correlator is reduced to a
computation in
an  $S_n$ orbifold CFT, rather than an $S_N$ orbifold CFT.

Once we have computed the correlators appearing on the rhs of
(\ref{Expand4ptN}) we
can perform the summation and obtain the desired 4-point
function appearing on the lhs for any values of $N$, $k$ and $l$. 
However,  using the global symmetry
under $SU(2)$ we can choose one of the four elements in the 
correlator to have a
$SU(2)$ spin `pointing up'; i.e., have $j_3=j$.  Thus it is 
sufficient to evaluate
$l=0$:
\be\label{DeflEq0}
\langle S_{N,0}(v)S^\dagger_{N-k,k}(w)\left((J_0^-)^p\sigma^P_n\right)(0)
\left((J_0^+)^q\sigma^{Q\dagger}_n\right)(a)\rangle
\ee
where the incoming state of the system has been taken to have `spin up'.

For the correlation function (\ref{DeflEq0}) the summation in
(\ref{Expand4ptN}) contains only one term ($b=0$), and we get:
\bea\label{TempLargeNAp}
&&\langle S_{N,0}(v)S^\dagger_{N-k,k}(w)\left((J_0^-)^p\sigma^P_n\right)(0)
\left((J_0^+)^q\sigma^{Q\dagger}_n\right)(a)\rangle_{S_N}
%\nonumber\\
%&&\qquad
=(v-w)^{-(N-n)/2}\left(\frac{(N-k)!n!}{N!(n-k)!}\right)^\frac{1}{2}\nonumber\\
&&\qquad\times\langle
{\tilde S}^{(1)}_{n,0}(v){\tilde S}^{(1)\dagger}_{n-k,k}(w)
\left((J_0^-)^p\sigma^P_n\right)(0)
\left((J_0^+)^q\sigma^{Q\dagger}_n\right)(a)\rangle_{S_n}
\eea

As discussed in section \ref{SectGen4ptRed}, every term in the right-hand side of the last expression can be reduced to the correlation function which has a form\footnote{This is accomplished by 
a  global $SU(2)$ rotation.} 
\be\label{genSSssA}
\langle S_{n-l,l}(v)S^\dagger_{n-k,k}(w)\sigma^P_n(0)
\left((J_0^+)^s\sigma^{Q\dagger}_n\right)(a)\rangle.
\ee
In the remaining part of this appendix we will derive the relation (\ref{SumRule}) which gives a further reduction of this correlator.

Since operator $J_0^+$ annihilates the highest weight state $\sigma^P_n$,
we can rewrite the correlation function (\ref{genSSssA}) in the following form:
\bea\label{SumRule1}
&&\langle (J_0^-)^s\left\{S_{n-l,l}(v)S^\dagger_{n-k,k}(w)\right\}\sigma^P_n(0)
\sigma^{Q\dagger}_n(a)\rangle\\
&&\qquad=\sum_{p=0}^s \frac{s!}{p!(s-p)!}
\langle \left\{(J_0^-)^{s-p} S_{n-l,l}(v)\right\}
\left\{(J_0^-)^{p} S^\dagger_{n-k,k}(w)\right\}\sigma^P_n(0)
\sigma^{Q\dagger}_n(a)\rangle.\nonumber
\eea
We should also note that the expressions in the curly brackets are
proportional to spin fields defined in (\ref{defGenerSpin}):
\bea
(J_0^-)^p S_{n-l,l}(z)&=&
(J_0^-)^p \sqrt{\frac{(n-l)!}{l!n!}}(J_0^-)^l
S_{n,0}(z)=
\sqrt{\frac{(n-l)!(p+l)!}{(n-l-p)!~l!}}S_{n-l-p,l+p}(z),\nonumber\\
(J_0^-)^p S^\dagger_{n-k,k}(z)&=&%(J_0^-)^p S^\dagger_{n-k,k}(z)=
\sqrt{\frac{k!(p+n-k)!}{(k-p)!(n-k)!}}S^\dagger_{n+p-k,k-p}(z).
\nonumber
\eea
Substituting these two expressions into (\ref{SumRule1}), we can obtain the
final result for (\ref{genSSssA}):
\bea\label{SumRuleApp}
&&\langle S_{n-l,l}(v)S^\dagger_{n-k,k}(w)\sigma^P_n(0)
\left((J_0^+)^s\sigma^{Q\dagger}_n\right)(a)\rangle\nonumber\\
&&\qquad=\sum_{p=0}^s \frac{s!}{p!(s-p)!}
\left(\frac{(n-k+p)!k!}{(n-k)!(k-p)!}
\frac{(n-l)!(s-p+l)!}{(n-l-s+p)!~l!}\right)^{1/2}\\
&&\qquad\quad\times
\langle S_{n-l-s+p,l+s-p}(v)
S^\dagger_{n-k+p,k-p}(w)
\sigma^P_n(0)\sigma^{Q\dagger}_n(a)\rangle\nonumber
\eea

\section{Four point function in terms of the cross ratio.}
\label{AppA}
\renewcommand{\theequation}{B.\arabic{equation}}
\setcounter{equation}{0}

In this paper we are interested in computing the correlator
\be\label{def4point}
\langle S_{N-l,l}(v)S^\dagger_{N-k,k}(w)
\left[(J_0^-)^p\sigma^{P}_n\right](0)
\left[(J_0^+)^q\sigma^{Q\dagger}_n\right](a)\rangle.
\ee
SL(2,C) invariance determines the 4-point correlator up to a function which depends only on one cross--ratio
\be\label{CrossRatioA}
x=\frac{v(w-a)}{w(v-a)}.
\ee
Let us write  the correlator (\ref{def4point}) in terms of a function
of this cross ratio. Using the general properties of the four point functions 
in two dimensional CFT (see for example \cite{CFT}), we get:
\bea\label{CrossTwistA}
&&\langle A_1(v)A_2(w)
\left[(J_0^-)^p\sigma^{P}_n\right](0)
\left[(J_0^+)^q\sigma^{Q\dagger}_n\right](a)\rangle=
v^{(h_2+Q-2P-2h_1)/3}w^{(h_1+Q-2P-2h_2)/3}\nonumber\\
&&\phantom{\frac{e^x}{f^y}}\times
(-a)^{(h_1+h_2-2P-2Q)/3}(v-w)^{(P+Q-2h_1-2h_2)/3}(v-a)^{(-2Q-2h_1+P+h_2)/3}\\
&&\phantom{\frac{e^x}{f^y}}\times(w-a)^{(-2Q-2h_2+P+h_1)/3}
f\left(\frac{v(w-a)}{w(v-a)}\right)\nonumber
\eea
In this expression $h_1$ and $h_2$ are conformal dimensions of $A_1$ and
$A_2$ and we also used the fact that by definition, the conformal dimension
of $\sigma_n^P$ is equal to $P$.

In later calculations it will be convenient to put the two twist
operators at $z=0$ and at $z=\infty$, so that the
branching points of the map from the $z$ space to the covering space
are at the `north' and `south' poles
of the sphere. Thus we should look at the limit 
$a\rightarrow\infty$. In this case the cross ratio (\ref{CrossRatioA}) becomes
\be
x\rightarrow\frac{v}{w}
\ee
and we can extract the function $f(x)$ from the limit:
\bea\label{DefFuncCrossR}
&&\lim_{a\rightarrow\infty}
\frac{\langle A_1(v)A_2(w)
\left[(J_0^-)^p\sigma^{P}_n\right](0)
\left[(J_0^+)^q\sigma^{Q\dagger}_n\right](a)\rangle}
{\langle
\sigma^Q_n(0)\sigma^{Q\dagger}_n(a)\rangle}=(-1)^{-P-Q}w^{Q-P-h_1-h_2}\\
&&\qquad\times\left(\frac{v}{w}\right)^{(h_2+Q-2P-2h_1)/3}
(\frac{v}{w}-1)^{(P+Q-2h_1-2h_2)/3}f\left(\frac{v}{w}\right)\nonumber
\eea

We will consider only the case when $A_1$ and $A_2$ are the spin operators. 
In this case it is convenient to introduce a more explicit notation for the 
function of the cross ratio $f(x)$: 
\bea
&&f(n,l,k|P,p;Q,q|x)\equiv
\langle S_{n-l,l}(v)S_{n-k,k}^\dagger(w)
\left[(J_0^-)^p\sigma^{P}_n\right](0)
\left[(J_0^+)^q\sigma^{Q\dagger}_n\right](a)\rangle\nonumber\\
&&\phantom{\frac{e^x}{f^y}}\times
v^{-(Q-2P-n/4)/3}w^{-(Q-2P-n/4)/3}(w-a)^{(2Q-P+n/4)/3}\nonumber\\
&&\phantom{\frac{e^x}{f^y}}\times
(-a)^{(2P+2Q-n/2)/3}(v-w)^{(n-P-Q)/3}(v-a)^{(2Q-P+n/4)/3}
\eea
Then equation (\ref{DefFuncCrossR}) for this case becomes:
\bea\label{ADefFuCrRSpec}
&&\lim_{a\rightarrow\infty}
\frac{\langle S_{n-l,l}(v)S_{n-k,k}^\dagger(w)
\left[(J_0^-)^p\sigma^{P}_n\right](0)
\left[(J_0^+)^q\sigma^{Q\dagger}_n\right](a)\rangle}
{\langle
\sigma^Q_n(0)\sigma^{Q\dagger}_n(a)\rangle}=w^{Q-P-n/2}\\
&&\qquad\times (-1)^{-P-Q}\left(\frac{v}{w}\right)^{(Q-2P-n/4)/3}
(\frac{v}{w}-1)^{(P+Q-n)/3}f\left(n,l,k|P,p;Q,q|\frac{v}{w}\right)\nonumber
\eea

\section{Computing OPE coefficients}
\label{AppThreePoint}
\renewcommand{\theequation}{C.\arabic{equation}}
\setcounter{equation}{0}

In this appendix we will compute the leading contributions to the three--point functions which appear in the expansion 
(\ref{ABCDrat3}) for the four--point function (\ref{zone}). In particular, we will concentrate on the last two channel depicted in figure \ref{figFusion}(b), where twist operator comes close to the spin operator. 

\subsection{Leading term in the OPE for $Q=\frac{n+1}{2}$.}
\label{SubsLeadPlus}

We begin with looking at an OPE of a spin operator
$S_{k,n-k}$ and a
twist  operator $\sigma_n^{\frac{n+1}{2}}$:
\be
S_{k,n-k}(w)\sigma_n^{\frac{n+1}{2}}(0)=
\sum_i C_{i}(n,k,\frac{n+1}{2})
w^{-\frac{n+1}{2}-\frac{n}{4}+\Delta_i}{\cal A}_i(0).
\ee
Here we have introduced a self-evident notation for the fusion coefficient.

As mentioned before, our general method of evaluating quantities in
the orbifold CFT is
to pass to the covering space, where we just get a $c=6$ CFT, and
operators in the $z$
plane map to appropriate operators in this latter CFT.  The operators
containing the twists
are located at $z=0, z=\infty$ (\ref{zone}).  Thus  we can go to the covering  space using the map:
\be\label{ztMap}
z=t^n.
\ee
The twist operator at the point $z=0$ corresponds to the following insertion
in the $t$ plane:
\be
\sigma^{Q}_n(0)~\rightarrow ~
{\hat\sigma}^{Q}_n(t=0)~=~:\exp\left(iQe_a\Phi^a(0)\right):.
\ee

The spin operator is located at $z=w$, but in the $t$ plane this point has $n$
different images\footnote{Note that $r$ is a complex number and not the absolute value of $t_j$.}:
\be
t_j=r\alpha_j,\qquad r^n=w,\qquad \alpha_j=\exp\left(\frac{2\pi ij}{n}\right),
\ee
thus the image of the spin operator in the $t$ plane is given by:
\bea
&&S_{k,n-k}(w)=\sqrt{\frac{(n-k)!}{k!n!}}\left(J^+_0\right)^k
S_{0,n}(w)\nonumber\\
&&\rightarrow\sqrt{\frac{(n-k)!}{k!n!}}\left({\hat J}^+_0\right)^k
\prod_{j=1}^n\left(\frac{dz}{dt}\right)_{t=r\alpha_j}^{-1/4}
:\exp\left(-\frac{i}{2}e_a\Phi^a(r\alpha_j)\right):
\eea
Here we introduced a notation for the modes of the current in the $t$ plane:
\be
{\hat J}^a_m=\oint\frac{dt}{2\pi i}t^m {\hat J}^a(t).
\ee
Due to the structure of the map (\ref{ztMap}), these operators are related to the modes in the $z$ plane in a simple way:
\be
{\hat J}^a_m=J^a_{m/n}.
\ee

Bringing the spin and twist operators together and writing the
exponentials as a single
normal ordered expression, we get:
\bea\label{form1p}
&&S_{k,n-k}(w)\sigma^{Q}_n(0)\rightarrow\sqrt{\frac{(n-k)!}{k!n!}}
r^{-Qn}r^{n(n-1)/4}n^{n/4}\left(nr^{n-1}\right)^{-n/4}\nonumber\\
&&\qquad\times\left({\hat J}^+_0\right)^k\left\{
:\exp\left(iQe_a\Phi^a(0)-\sum_{j=1}^n
\frac{i}{2}e_a\Phi^a(r\alpha_j)\right):\right\}
\eea
Then after writing out the current operators explicitly as contour integrals,
we can extract the leading power of $w=r^n$ for $Q=\frac{n+1}{2}$:
\bea\label{3ptLeadp}
&&S_{k,n-k}(w)\sigma^{\frac{n+1}{2}}_n(0)\rightarrow\nonumber\\
&&\qquad\sqrt{\frac{(n-k)!}{k!n!}}r^{-\frac{n(n+1)}{2}+kn}
\left\{\prod_{m=1}^k\oint\frac{dt_m}{2\pi i}\right\}
\left\{\prod_m^kt^{1-n}_m\right\}
\left\{\prod_{l<m}^k(t_l-t_m)^2\right\}\nonumber\\
&&\qquad\times
\left[:\exp\left(\frac{i}{2}e_a\Phi^a(0)+
ie_a\sum_m^k\Phi^a(t_m)\right):\right]+
O(w^{-\frac{n+1}{2}+k+1})
\eea

We now observe that the contour integrals in the above expression are
precisely those
that correspond to an application of $k$ modes of ${\hat J}^+_{-n}$:
\bea\label{3ptLead}
&&S_{k,n-k}(w)\sigma^{\frac{n+1}{2}}_n(0)\rightarrow\nonumber\\
&&\quad\sqrt{\frac{(n-k)!}{k!n!}}r^{-\frac{n(n+1)}{2}+kn}
\left({\hat J}^+_{-n}\right)^k
\left[:\exp\left(\frac{i}{2}e_a\Phi^a(0)
\right):\right]+
O(w^{-\frac{n+1}{2}+k+1})
\eea

To evaluate the fusion coefficient for this case (we call it
$C_{i}(n,k,\frac{n+1}{2})$)
we need to rewrite the right hand side of the last expression in terms of
the normalized operator ${\cal A}$. In other words, we have to calculate the
norm of the state
\be\label{mergeNN}
\left({\hat J}^+_{-n}\right)^k
\left[:\exp\left(\frac{i}{2}e_a\Phi^a(0)
\right):\right]|0\rangle_{NS}
\ee
corresponding to the operator appearing on the right hand side of
(\ref{3ptLead}).

We will use the $SU(2)$ algebra in order to evaluate the norm of the
state. Let  us label
states by their $SU(2)$ quantum numbers $|j,m\rangle$. In this notation the
state $:\exp\left(\frac{i}{2}e_a\Phi^a(0)\right):|0\rangle_{NS}$ can
be written as
\be
|\frac{1}{2},\frac{1}{2}\rangle.
\ee
and the state (\ref{mergeNN}) becomes
\be\label{StateNNSU2}
\left({\hat J}^+_{-n}\right)^k
|\frac{1}{2},\frac{1}{2}\rangle
\ee
Then using the standard manipulation with $SU(2)$ algebra, we find:
\be
\langle \frac{1}{2},\frac{1}{2}|\left({\hat J}^-_{n}\right)^k
\left({\hat J}^+_{-n}\right)^k
|\frac{1}{2},\frac{1}{2}\rangle=\frac{k!(n-1)!}{(n-k-1)!}.
\ee
\be
F(n,k)=\frac{k!(n-1)!}{(n-k-1)!}
\ee
This allows us to rewrite the operator corresponding to the state
(\ref{mergeNN}) in terms
of a  normalized operator ${\cal A}_{n,k}$:
\be\label{Final+OPE}
S_{k,n-k}(w)\sigma^{\frac{n+1}{2}}_n(0)=\left(\frac{n-k}{n}\right)^{1/2}
w^{-\frac{n+1}{2}+k}{\cal A}_{n,k}(0)+
O(w^{-\frac{n+1}{2}+k+1})
\ee
Thus the  the dimension of the leading operator ${\cal A}_{n,k}$ in the OPE is:
\be\label{Lead+Dim}
\Delta_{n,k}=k+\frac{n}{4}
\ee
and the fusion coefficient  is
\be\label{Lead+C}
C(n,k,\frac{n+1}{2})=\left(\frac{n-k}{n}\right)^{1/2}.
\ee
We will also need the $SU(2)$ quantum numbers of the state ${\cal A}_{n,k}$:
\be
j_3=k+\frac{1}{2},\qquad j=k+\frac{1}{2}.
\ee
They can be found by applying $J^3_0$ and $J^2=J^-_0J^+_0+2J_0^3+(J_0^3)^2$ to
the state (\ref{StateNNSU2}).

\subsection{Leading term in the OPE for $Q=\frac{n-1}{2}$.}

Let us now evaluate the leading term in the following OPE:
\be\label{MinusOPE}
S_{k,n-k}(w)\sigma^{\frac{n-1}{2}}_n(0)=\sum_i C_i(n,k,\frac{n-1}{2})
w^{-\frac{n-1}{2}-\frac{n}{4}+\Delta_i}{\cal A}_i(0)
\ee
Unfortunately, in this case the direct extraction of the leading power from 
the analog of (\ref{form1p}) is more complicated, and we will find it 
convenient to proceed in a somewhat different manner.

Let us look at the images of $\sigma_n^{\frac{n\pm 1}{2}}$ on the covering
space:
\be
\label{sigma-J+}
{\hat \sigma}_n^{\frac{n-1}{2}}(0)=
J_{n}^{-}{\hat \sigma}_n^{\frac{n+1}{2}}(0).
\ee
Substituting this into the left hand side of (\ref{MinusOPE}), we get
\bea\label{OPE+1}
&&{\hat\sigma}^{\frac{n-1}{2}}_n(0){\hat S}_{k,n-k}(r)=
\oint_0 \frac{dt}{2\pi i}t^nJ^{-}(t)
{\hat\sigma}^{\frac{n+1}{2}}_n(0){\hat S}_{k,n-k}(r)\nonumber\\
&&=J^{-}_{n}\left[{\hat\sigma}^{\frac{n-1}{2}}_n(0){\hat S}_{k,n-k}(r)\right]-
{\hat\sigma}^{\frac{n+1}{2}}_n(0)\oint_{w^{1/n}}
\frac{dt}{2\pi i}t^nJ^{-}(t){\hat S}_{k,n-k}(r)
\eea
Here $\oint_{w^{1/n}}$ means integration over the contour which goes around
all images of the point $z=w$.

It will be helpful to look at the last expression in terms of
operators on the $z$ space:
\be\label{aug1}
\oint_{w^{1/n}}\frac{dt}{2\pi i}t^nJ^{-}(t){\hat S}_{k,n-k}(r)\rightarrow
\oint_{w}\frac{dz}{2\pi i}z J^{-}(z)S_{k,n-k}(w)=wJ^{(z)-}_0 S_{k,n-k}(w)
\ee
The last obtained form is useful because the definitions of the
$S_{k, n-k}$ were already in
a form where operators $J^{(z)-}_0$ acted upon $S_{0,n}$. Recalling
the definition
\be
S_{k,n-k}(w)=S^\dagger_{n-k,k}(w)=
=\sqrt{\frac{k!}{(n-k)!n!}}\left(J^{(z)-}_0\right)^{n-k}S_{n,0}(w),
\ee
we can rewrite the rhs of (\ref{aug1}) as
\be
wJ^{(z)-}_0 S_{k,n-k}(w)=w\sqrt{k(n-k+1)}S_{k-1,n-k+1}(w)
\ee
Substituting this into the OPE (\ref{OPE+1}) one gets:
\bea\label{OPE+2}
{\hat S}_{k,n-k}(r){\hat\sigma}^{\frac{n-1}{2}}_n(0)
=J^{-}_{n}\left[{\hat\sigma}^{\frac{n+1}{2}}_n(0){\hat S}_{k,n-k}(r)\right]-
r^n\sqrt{k(n-k+1)}
{\hat\sigma}^{\frac{n+1}{2}}_n(0){\hat S}_{k-1,n-k+1}(r)\nonumber
\eea
We can now use the fusion rule for $S_{k, n-k}$ with $\sigma_n^{n+1\over 2}$
which has already been found in subsection \ref{SubsLeadPlus}. Then the
above equation gives
\bea\label{OPE+3}
&&{\hat S}_{k,n-k}(r){\hat\sigma}^{\frac{n-1}{2}}_n(0)=
%\left(\frac{(n-k)!}{n!k!}\right)^{1/2}r^{-\frac{n(n+1)}{2}+kn}J^{-}_{n}
%\left(J^{+}_{-n}\right)^k|\frac{1}{2},\frac{1}{2}\rangle\nonumber\\
%&&\qquad -\sqrt{k(n-k+1)}\left(\frac{(n-k+1)!}{n!(k-1)!}\right)^{1/2}
%r^{n-\frac{n(n+1)}{2}+(k-1)n}
%\left(J^{+}_{-n}\right)^{k-1}|\frac{1}{2},\frac{1}{2}\rangle\nonumber\\
%&&=\left(\frac{(n-k)!}{n!k!}\right)^{1/2}r^{-\frac{n(n+1)}{2}+kn}
%\left(J^{+}_{-n}\right)^{k-1}|\frac{1}{2},\frac{1}{2}\rangle
%\left(-k(n-k+1)+\sum_{j=0}^{k-1}\left[n-2(\frac{1}{2}+j)\right]\right)
%\nonumber\\
%&&=
-k\left(\frac{(n-k)!}{n!k!}\right)^{1/2}r^{-\frac{n(n+1)}{2}+kn}
\left(J^{+}_{-n}\right)^{k-1}|\frac{1}{2},\frac{1}{2}\rangle
\eea
where in the second step we have used the algebra of SU(2).

We see that we generate the same operators at leading order in the
OPE as the ones that
appeared for the case of $\sigma_n^{n+1\over 2}$.  The norm of the
state on the rhs  of
(\ref{OPE+3}) was evaluated  in  subsection \ref{SubsLeadPlus}, so for the
OPE (\ref{MinusOPE}) we
get
\bea\label{Final-OPE}
S_{k,n-k}(w)\sigma^{\frac{n-1}{2}}_n(0)&\approx&C(n,k,\frac{n-1}{2})
w^{-\frac{n-1}{2}-\frac{n}{4}+(\frac{n}{4}-1+k)}{\cal A}_{n, k-1}(0)
\eea

Thus the dimension of the leading operator in the OPE is
\be
\Delta_{n, k-1}=\frac{n}{4}-1+k
\ee
and the fusion coefficient is
\bea\label{LaedC}
C(n,k,\frac{n-1}{2})=-k\left(\frac{(n-k)!}{n!k!}\right)^{1/2}
\left(\frac{(k-1)!(n-1)!}{(n-k)!}\right)^{1/2}=-\left(\frac{k}{n}\right)^{1/2}.
\eea

\section{Correlation function where one twist operator has $j_3=-j$.}
\label{SectChirTwist}
\renewcommand{\theequation}{D.\arabic{equation}}
\setcounter{equation}{0}

Our main goal in this paper is to evaluate the four point function involving
two twist operators and two spin operators:
\be
\langle S_{N-l,l}(v)
S_{N-k,k}^\dagger(w)\left((J_0^-)^p\sigma^P_n\right)(0)
\left((J_0^+)^q\sigma^{Q\dagger}_n\right)(a)\rangle.
\ee
In section \ref{SectGen4ptRed} we have shown that evaluation of such
correlation functions for the $S_N$ orbifold can be reduced to the
calculation of the
correlators:
\be\label{PattCorFunc}
\langle S_{n-l,l}(v)S^\dagger_{n-k,k}(w)\left((J_0^-)^s\sigma^P_n\right)(w)
\sigma^{Q\dagger}_n(a+w)\rangle.
\ee
for an  $S_n$ orbifold. Evaluation of this expression in turn can be reduced to
calculation of two ``basic correlators'' and some
combinatoric factors. In  section \ref{SectBasic} we have presented the
expressions
for ``basic correlators'' and here we will collect all this information
together and derive the result for (\ref{PattCorFunc}).
As before it will be convenient to consider the cases of $P=Q$ and $P\ne Q$
separately.

\subsection{Four point function with $P=Q$.}

We begin with evaluation of the four point function:
\be
\langle S_{n-l,l}(v)S^\dagger_{n-k,k}(w)\sigma^P_n(0)
\left((J_0^+)^s\sigma^{P\dagger}_n\right)(a)\rangle,
\ee
where
$P$ takes the two possible values $\frac{1}{2}(n\pm 1)$. Let us
recall the expression for
this four  point function in terms of basic correlators (\ref{SumRule}):
\bea\label{HomSR1}
&&\langle S_{n-l,l}(v)S^\dagger_{n-k,k}(w)
\left((J_0^-)^s\sigma^P_n\right)(0)
\sigma^{P\dagger}_n(a)\rangle\nonumber\\
&&=\sum_{p=0}^s \frac{s!}{p!(s-p)!}
\left(\frac{(n-k+p)!k!}{(n-k)!(k-p)!}
\frac{(n-l)!(s-p+l)!}{(n-l-s+p)!~l!}\right)^{1/2}(-1)^s\\
&&\qquad\quad\times
\langle S_{n-l-s+p,l+s-p}(v)
S^\dagger_{n-k+p,k-p}(w)
\sigma^P_n(0)\sigma^{P\dagger}_n(a)\rangle\\
&&=\left(\frac{k!(n-l)!}{l!(n-k)!}\right)^{1/2}
\sum_{p=0}^{k-l} \frac{(-1)^{k-l}(k-l)!}{p!(k-l-p)!}
\langle S_{n-k+p,k-p}(v)S^\dagger_{n-k+p,k-p}(w)
\sigma^P_n(0)\sigma^{P\dagger}_n(a)\rangle\nonumber
\eea
At the last step we used the charge conservation which implies the relation 
between $k,l$ and $s$: $s=k-l$.

Since an application of  $J_0^+$ does not change the dimension of an
operator,  we obtain
the same overall factors on both sides of this equation when
expressing the correlators
through functions of the cross ratio; thus we can compare functions $f$
\bea\label{HomSR2}
&&f(n,l,k|P,k-l;P,0|x)=
\left(\frac{k!(n-l)!}{l!(n-k)!}\right)^{1/2}(-1)^{k-l}\nonumber\\
&&\qquad\times
\sum_{p=0}^{k-l} \frac{(k-l)!}{p!(k-l-p)!}f(n,k-p,k-p|P,0;P,0|x)
\eea

First take $P=\frac{n+1}{2}$. In this
case the correlation functions entering the rhs of (\ref{HomSR2})
are given by (\ref{FFact++}), and evaluating the sum in (\ref{HomSR2}), we get:
\bea\label{HomSRRes1}
&&f(n,l,k|\frac{n+1}{2},k-l;\frac{n+1}{2},0|x)=
\frac{1}{n}\left(\frac{k!(n-l)!}{l!(n-k)!}\right)^{1/2}
(x-1)^{-n/2-1/3}x^{3n/4-1/3-k}\nonumber\\
&&\qquad\times
(1+x)^{k-l}\left\{(n-l)x+2l-k+\frac{2(k-l)}{1+x}\right\}(-1)^{k-l+n+1}.
\eea
The calculations for the case of $P=(n-1)/2$ can be done in the same way
(one has to use (\ref{FFact--}) instead of (\ref{FFact++})) and we get
\bea\label{HomSRRes2}
&&f(n,l,k|\frac{n-1}{2},k-l;\frac{n-1}{2},0|x)=
\frac{1}{n}\left(\frac{k!(n-l)!}{l!(n-k)!}\right)^{1/2}
(x-1)^{-n/2+1/3}x^{3n/4-2/3-k}
\nonumber\\
&&\qquad\times
(1+x)^{k-l}\left\{lx+n+k-2l+\frac{2(l-k)}{1+x}\right\}(-1)^{k-l+n+1}.
\eea

\subsection{Four point function with $P\ne Q$.}

We now address  the case $P\ne Q$.
We look at  the correlator
\be
\langle S_{n-l,l}(v)S^\dagger_{n-k,k}(w)
\left((J_0^-)^s\sigma^{\frac{n-1}{2}}_n\right)(0)
\sigma^{\frac{n+1}{2}\dagger}_n(a)\rangle.
\ee
The expansion of this four point function in terms of basic correlators is
given by (\ref{SumRule}), and the analog of (\ref{HomSR2}) reads:
\bea\label{HetSR2}
&&
f(n,l,k|\frac{n-1}{2},k-l-1;\frac{n+1}{2},0|x)=
\left(\frac{k!(n-l)!}{l!(n-k)!}\right)^{1/2}(-1)^{k-l-1}
\\
&&\qquad\times\sum_{p=0}^{k-l-1}
\left\{\frac{(k-l-1)!}{p!(k-l-1-p)!}
\frac{1}{\sqrt{(n-k+p+1)(k-p)}}\right.
\nonumber\\
&&\qquad\left.\phantom{\frac{ty}{ty}}\times
f(n,k-p-1,k-p|\frac{n-1}{2},0;\frac{n+1}{2},0|x)\right\}
\nonumber
\eea
To evaluate this correlator we use the expression for the basic correlation
function (\ref{FFunc+-}), then one gets:
\bea\label{HetSRRes1}
&&f(n,l,k|\frac{n-1}{2},k-l-1;\frac{n+1}{2},0|x)\nonumber\\
&&=(-1)^{k-l+n}\frac{1}{n}\left(\frac{k!(n-l)!}{l!(n-k)!}\right)^{1/2}
x^{3n/4-k}(x-1)^{-n/2+1}(1+x)^{k-l-1}
\eea

\section{General correlation function.}
\label{SectGenCorr}
\renewcommand{\theequation}{E.\arabic{equation}}
\setcounter{equation}{0}

\subsection{Method of calculation.}
In this section we perform the last step in evaluating the general four point
function:
\be\label{defGen4pt}
\langle S_{n-l,l}(v)S^\dagger_{n-k,k}(w)
\left((J_0^-)^p\sigma^{P}_n\right)(0)
\left((J_0^+)^q\sigma^{Q\dagger}_n\right)(a)\rangle.
\ee
As before we will be interested in the reduced four point function, which
depends only upon the cross ratio:
\be\label{zthree}
f(n,l,k|P,p;Q,q|x),
\ee
and the complete function (\ref{defGen4pt}) can be recovered using
(\ref{DefFIndFin}).
Since the dimension of $S_{n-k,k}$ does not depend on $k$ and the dimension of
$(J_0^-)^p\sigma^{P}_n$ is independent on $p$, the prefactor in front of $f$
is universal; it depends only upon $n, P$ and $Q$. Thus all relations between
various correlators of the type (\ref{defGen4pt}) which have the same
values of $n, P$
and
$Q$ can  be rewritten in terms of reduced functions $f$.

Let us start by moving the contours of $J_0^+$ away from the point $z=a$ in
(\ref{defGen4pt}). Then we get:
\bea\label{4ptmoveJ}
&&\left< S_{n-l,l}(v)S^\dagger_{n-k,k}(w)
\left((J_0^-)^p\sigma^{P}_n\right)(0)
\left((J_0^+)^q\sigma^{Q\dagger}_n\right)(a)\right>\\
&&\qquad=
(-1)^q\sum_{a=0}^q \frac{q!}{a!(q-c)!}
\left< (J_0^+)^a\left[S_{n-l,l}(v)S^\dagger_{n-k,k}(w)\right]
\left((J_0^+)^{q-a}(J_0^-)^p\sigma^{P}_n\right)(0)
\sigma^{Q\dagger}_n(a)\right>\nonumber
\eea
Let us consider the operator
\be\label{calM}
{\cal M}(a,b,P)\equiv (J_0^+)^{a}(J_0^-)^b\sigma^{P}_n(0).
\ee
Since $\sigma^{P}_n$ is the highest member of an $SU(2)$ multiplet, it is
annihilated by $J_0^+$. Then using standard manipulations, we get for $b>a$:
\be
{\cal M}(a,b,P)=\frac{b!}{(b-a)!}\frac{(2P-b+a)!}{(2P-b)!}
(J_0^-)^{b-a}\sigma^{P}_n(0)
\ee
Substituting this in (\ref{4ptmoveJ}), one gets:
\bea\label{4ptmoveJ1}
&&\left< S_{n-l,l}(v)S^\dagger_{n-k,k}(w)
\left((J_0^-)^p\sigma^{P}_n\right)(0)
\left((J_0^+)^q\sigma^{Q\dagger}_n\right)(a)\right>\nonumber\\
&&\qquad=
(-1)^q\sum_{a=0}^q \frac{q!}{a!(q-c)!}\frac{p!(2P-p+q-a)!}{(p-q+a)!
(2P-p)!}
\\
&&\qquad\qquad\times
\left< (J_0^+)^a\left[S_{n-l,l}(v)S^\dagger_{n-k,k}(w)\right]
\left((J_0^-)^{p-q+a}\sigma^{P}_n\right)(0)
\sigma^{Q\dagger}_n(a)\right>.\nonumber
\eea

We can distribute $(J_0^+)^a$ between the two spin operators and use the 
charge conservation:
\be\label{ChargeConserv}
k-l-p+P+q-Q=0.
\ee
to get the four point functions in (\ref{4ptmoveJ1}) in terms of
functions of the cross ratio:
\bea
\label{4ptmoveJ3}
&&f(n,l,k|P,p;Q,q|x)
=(-1)^q\left(\frac{(n-k)!l!}{k!(n-l)!}\right)^{1/2}\frac{p!}{(2P-p)!}
\nonumber\\
&&\qquad\times\sum_{c=0}^q \frac{q!}{c!(q-c)!}\frac{(2P-p+q-c)!}{(p-q+c)!}
\sum_{b=0}^c\left(\frac{(k+b)!}{(n-k-b)!}
\frac{(n-l+c-b)!}{(l-c+b)!}
\right)^{1/2}
\nonumber\\
&&\qquad\phantom{\frac{e^x}{d^s}}\times \frac{c!}{b!(c-b)!}
f(n,l-c+b,k+b|P,p-q+c;Q,0|x)
\eea
The four point functions appearing on the rhs of (\ref{4ptmoveJ3})
should be taken from the equations (\ref{HomSRRes1}),(\ref{HomSRRes2}) or
(\ref{HetSRRes1}) depending
upon the values of $P$ and $Q$. We consider these three cases separately.

\subsection{Calculation for $P=Q=\frac{n+1}{2}$.}
\label{AppE2}

Substituting the expression (\ref{HomSRRes1}) into (\ref{4ptmoveJ3}), one
gets
\bea\label{SumHomog1}
&&f(n,l,k|\frac{n+1}{2},p;\frac{n+1}{2},q|x)
=(-1)^q\left(\frac{(n-k)!l!}{k!(n-l)!}\right)^{1/2}\frac{p!}{(n+1-p)!}
\nonumber\\
&&\qquad\times\frac{1}{n}(x-1)^{-n/2-1/3}x^{3n/4-1/3-k}
(1+x)^{k-l}(-1)^{k-l+n+1}\nonumber\\
&&\qquad\times\sum_{c=0}^q \frac{q!}{c!(q-c)!}\frac{(n+1-p+q-c)!}{(p-q+c)!}
\sum_{b=0}^c\frac{(k+b)!}{(n-k-b)!}
\frac{(n-l+c-b)!}{(l-c+b)!}\frac{c!}{b!(c-b)!}\nonumber\\
&&\qquad\phantom{\frac{e^x}{d^s}}
\times x^{-b}(-1-x)^c\left\{\alpha+\beta c+\gamma b\right\}
\eea
Here we introduced the convenient notation:
\bea\label{defAlBeGa}
\alpha&=&(n-l)x+2l-k+2\frac{k-l}{1+x}\nonumber\\
\beta&=&x-2+\frac{2}{1+x}\nonumber\\
\gamma&=&1-x
\eea

We will not simplify the above result further for general values of
the paramaters. But we
had shown in section \ref{SectGen4ptRed} that to evaluate the four point
functions
(\ref{DeflEq0}) a special
subset of these correlators was needed. We will now obtain an
explicit expression for these
special correlators.

For this special subset we have $l=0$.
Then by charge conservation $p=k+q$ and we get
\bea\label{resHom1N}
&&f(n,0,k|\frac{n+1}{2},p;\frac{n+1}{2},q|x)
=(-1)^{q}
\left(\frac{(n-k)!l!}{k!(n-l)!}\right)^{1/2}\frac{(k+q)!}{(n+1-k-q)!}
\nonumber\\
&&\qquad\times\frac{(-1)^{k+n+1}}{n}(x-1)^{-n/2-1/3}x^{3n/4-1/3-k}
(1+x)^{k}n!\nonumber\\
&&\qquad\times\sum_{c=0}^q \frac{q!}{c!(q-c)!}(n+1-k-c)
\left(-\frac{1+x}{x}\right)^c\left\{\alpha+(\beta+\gamma) c\right\}
\eea
One can rewrite the last line in this expression in the following form
\be
\left.\left(n+1-k-y\frac{\partial}{\partial y}\right)
\left\{\alpha+(\beta+\gamma) y\frac{\partial}{\partial y}\right\}
(1+y)^q\right|_{y=-\frac{1+x}{x}},
\ee
After substituting the values of $\alpha$, $\beta$ and $\gamma$ from
(\ref{defAlBeGa}), the equation (\ref{resHom1N}) reads
\bea\label{FinRes++2}
&&f(n,0,k|\frac{n+1}{2},p;\frac{n+1}{2},q|x)
=(-1)^q\left(\frac{(n-k)!}{k!n!}\right)^{1/2}\frac{(k+q)!}{(n+1-k-q)!}
\nonumber\\
&&\qquad\times\frac{(-1)^{k+n+1}}{n}(x-1)^{-n/2-1/3}x^{3n/4-1/3-k}
(1+x)^{k}n!\nonumber\\
&&\qquad\times
\left.\left(n+1-k-y\frac{\partial}{\partial y}\right)
\left\{nx+k\frac{1-x}{1+x}+\frac{1-x}{1+x}y\frac{\partial}{\partial y}\right\}
(1+y)^q\right|_{y=-\frac{1+x}{x}}
\eea

\subsection{Calculation for $P=Q=\frac{n-1}{2}$.}

Substituting the expression (\ref{HomSRRes2}) into (\ref{4ptmoveJ3}), one
gets for $l=0$:
\bea\label{resHom2N}
&&f(n,0,k|\frac{n-1}{2},p;\frac{n-1}{2},q|x)
=(-1)^q\left(\frac{(n-k)!}{k!n!}\right)^{1/2}\frac{(k+q)!}{(n-1-k-q)!}
\nonumber\\
&&\qquad\times\frac{(-1)^{k+n+1}}{n}(x-1)^{-n/2+1/3}x^{3n/4-2/3-k}
(1+x)^{k}n!\nonumber\\
&&\qquad\times\sum_{c=0}^q \frac{q!}{c!(q-c)!}\frac{1}{n-k-c}
\left(-\frac{1+x}{x}\right)^c\left\{\tilde\alpha+
(\tilde\beta+\tilde\gamma) c\right\}
\eea
Here
\bea\label{DefTildAlbeGa}
\tilde\alpha=n+k-2\frac{k}{1+x},\qquad
\tilde\beta=-x+2-\frac{2}{1+x},\qquad%\nonumber\\
\tilde\gamma=x-1
\eea
The sum in the last line of (\ref{resHom2N}) can be rewritten in terms of the 
hypergeometric function:
\bea
&&\sum_{c=0}^q \frac{q!}{c!(q-c)!}\frac{1}{n-k-c}
\left(-\frac{1+x}{x}\right)^c\left\{\tilde\alpha+
(\tilde\beta+\tilde\gamma) c\right\}\nonumber\\
&&\qquad\left.=\frac{1}{n-k}
\left\{\tilde\alpha+(\tilde\beta+\tilde\gamma)
y\frac{\partial}{\partial y}\right\}
F(-q,k-n;k-n+1;y)\right|_{y=-\frac{1+x}{x}},
\eea
which leads to the final answer for this case:
\bea\label{FinRes--1}
&&f(n,0,k|\frac{n-1}{2},p;\frac{n-1}{2},q|x)
=(-1)^q\left(\frac{(n-k)!}{k!n!}\right)^{1/2}\frac{(k+q)!}{(n-1-k-q)!}
\nonumber\\
&&\qquad\times\frac{(-1)^{k+n+1}}{n}(x-1)^{-n/2+1/3}x^{3n/4-2/3-k}
(1+x)^{k}n!\frac{1}{n-k}\nonumber\\
&&\qquad\times
\left.\left\{n+k\frac{x-1}{x+1}+\frac{x-1}{x+1}
y\frac{\partial}{\partial y}\right\}
F(-q,k-n;k-n+1;y)\right|_{y=-\frac{1+x}{x}}
\eea

\subsection{Calculation for $P=\frac{n+1}{2}$, $Q=\frac{n-1}{2}$.}

Substituting the expression (\ref{HetSRRes1}) into (\ref{4ptmoveJ3}), one
gets for $l=0$:
\bea\label{SumHeter2}
&&f(n,0,k|\frac{n-1}{2},p;\frac{n+1}{2},q|x)
=(-1)^q\left(\frac{(n-k)!}{k!n!}\right)^{1/2}\frac{p!}{(n-1-p)!}
\nonumber\\
&&\qquad\times\frac{(-1)^{n+k}}{n}(x-1)^{-n/2+1}x^{3n/4-k}
(1+x)^{k-1}\nonumber\\
&&\qquad\times\sum_{c=0}^q \frac{q!n!}{c!(q-c)!}\frac{(n-1-p+q-c)!}{(p-q+c)!}
\frac{(k+c)!}{(n-k-c)!}
\left(-\frac{1+x}{x}\right)^c
\eea
Charge conservation leads to the relation
\be
p=k+q+P-Q=k+q-1,
\ee
which allows us to simplify the sum in (\ref{SumHeter2}):
\bea\label{FinRes+-}
&&f(n,0,k|\frac{n-1}{2},p;\frac{n+1}{2},q|x)
=(-1)^{q+k+n}\left(\frac{(n-k)!}{k!n!}\right)^{1/2}\frac{p!(n-1)!}{(n-1-p)!}
\nonumber\\
&&\qquad\times (x-1)^{-n/2+1}x^{3n/4-k}
(1+x)^{k-1}
%\nonumber\\
%&&\qquad
%\times
\left.\left(k+y\frac{\d}{\d y}\right)
\left(1+y\right)^q\right|_{y=-\frac{1+x}{x}}
\eea

\section{Large $N$ scaling of the correlators}
\label{AppLrgN}
\renewcommand{\theequation}{F.\arabic{equation}}
\setcounter{equation}{0}

In this paper we evaluated a large class of CFT amplitudes and found that transitions modifying the state of the black hole are suppressed by powers of $N$. Assuming that the final Ramond vacuum is a rotation of the initial state, we found that rotation by $k$ units leads to suppression by $1/N^k$, and in this appendix we 
will extend this result to more general transitions between vacua. 

Specifically, we consider the following correlation function 
\bea\label{Tcorrel}
\left\langle S_{N,0}(v)T^\dagger(w)\left((J_0^-)^p\sigma_n^P\right)(0)
\left((J_0^+)^q\sigma_n^{Q\dagger}\right)(a)\right
\rangle_{S_N}
\eea
We assumed that all spins in the initial state are pointing up, the twist operators interchange the first $n$ copies, and they flip $k$ spins. In this paper the final state $T$ was taken to be a rotation of the initial state, and one can consider two generalizations:
\begin{enumerate}
\item{Keeping the assumption that the product of $\sigma_n$ and $\sigma_n^\dagger$ corresponds to the trivial permutation, we conclude that $T$ must factorize
\bea\label{TandTilde}
T=c{\tilde S}^{(2)}_{N-n}{\tilde T}
\eea
However we now allow $T$ to belong to the representation with spin $j<N/2$, then ${\tilde T}$ belongs to the representation with $j'=j-\frac{N-n}{2}$. As before, the change of $j_3$ caused by the twist operators will be denoted by $k$ (i.e., the spin operator $T$ would have $j_3=\frac{N}{2}-k$), and such change can only happen in the first $n$ copies. The standard formulas for addition of angular momenta lead to inequalities
\bea\label{BoundsJk}
\frac{n}{2}\ge j'\ge \frac{n}{2}-k\quad\Rightarrow\quad 
\frac{N}{2}\ge j\ge \frac{N}{2}-k
\eea
Similar relations hold in the anti--holomorphic sector.
}
\item{More general correlators arise if one drops the assumption that the product $\sigma_n \sigma_n^\dagger$ gives a trivial permutation. In this case $T$ would contain operators ${\cal A}$ discussed in sections \ref{SectIntOPE}, \ref{SubsInterm}. Evaluation of these correlators involves nontrivial three--point functions in the orbifold CFT, and we leave it for future work. 
}
\end{enumerate}
In this appendix we explore the first option focusing on large values of $N$ (we assume that $N$ is much larger than $k$, ${\bar k}$). Then we will demonstrate that operator $T$ must have $j={\bar j}$ and 
\bea\label{FinConj}
\left\langle S_{N,0}(v)T_{j;j}^\dagger(w)\left((J_0^-)^p
({\bar J}_0^-)^{\bar p}\sigma_n^P\right)(0)
\left((J_0^+)^q
({\bar J}_0^+)^{\bar q}\sigma_n^{Q\dagger}\right)(a)
\right\rangle_{S_N}\sim N^{-(k+{\bar k})/2}
N^{(\frac{N}{2}-j)/2}
\eea
This expression reduces to (\ref{LrgNmain}) for $j=\frac{N}{2}$ and $k={\bar k}$. 

To prove (\ref{FinConj}), we will begin with justifying it for the highest member of the multiplet with given $j$, ${\bar j}$. According to (\ref{BoundsJk}) such correlator has $k=\frac{N}{2}-j$, ${\bar k}=\frac{N}{2}-{\bar j}$. Other values of $k$ and ${\bar k}$ will be considered in section \ref{SecLrgN2}.

\subsection{Highest weight states}

Operators $T$ corresponding to untwisted final states
can be written as superpositions of various spins, as in (\ref{defSpecSpin}), (\ref{defGenerSpin}). It is convenient to denote spins pointing up and down by arrows, then 
(\ref{defSpecSpin}) can be represented as
\bea\label{UpSpinAr}
S_{N,0} \rightarrow |\ua\dots\ua\rangle\equiv |\ua^N\rangle
\equiv |j=\frac{N}{2},j_3=\frac{N}{2}\rangle
\eea
Application of the twist operators to (\ref{UpSpinAr}), as in (\ref{Tcorrel}), produces a linear combinations of states containing various orientations of spins\footnote{We use a shorthand notation $\uparrow^n$ for $n$ consecutive arrows pointing up, there is also slot-by-slot correspondence between the left and right spins of a particular string.} 
\bea\label{SigmaToSpin}
|\downarrow^k \uparrow^{N-k};\downarrow^m
\uparrow^{k-m}\downarrow^{{\bar k}-m}
\uparrow^{N-k-{\bar k}+m}\rangle
\eea 
and all such terms appear with coefficients of order one. To evaluate the four point function (\ref{Tcorrel}), one should take a projection of (\ref{SigmaToSpin}) onto a final state $T$. In this paper we mainly focused on $T$ which was either the highest weight state (\ref{UpSpinAr}) with $j=\frac{N}{2}$ or some other state in the same representation
(i.e., $S_{N-k,k}$ defined by (\ref{defGenerSpin})). 

Now we will take $T$ to be the highest weight state 
$|j,j;{\bar j},{\bar j}\rangle$, with angular momenta $j$ and ${\bar j}$ in the left and in the right sectors. First we construct the holomorphic part of this state using the Young diagram for representation with angular momentum $j$:
\bea\label{JJsum}
|j,j\rangle_L=\frac{1}{\sqrt{s}}\left[|\uparrow^{\frac{N}{2}+j}\da^{\frac{N}{2}-j}\rangle_L+perm\right],
\eea
where permutations correspond to the Young diagram with $\frac{N}{2}+j$ boxes in the first row and $\frac{N}{2}-j$ boxes in the second row, and $s$ is the number of terms in the sum. Notice that we are focusing on $N\gg k$, then inequality (\ref{BoundsJk}) implies that coefficients in front of individual states in the right hand side of (\ref{JJsum}) are of order one (i.e., $s\sim N^0$). 

To construct the spin operator $T$, one should combine (\ref{JJsum}) with corresponding expression in the anti--holomorphic sector and sum over the orbit of the permutation group ($d$ is the dimension of the orbit):
\bea\label{OrbYoung}
&&|j,j;{\bar j},{\bar j}\rangle=\frac{1}{\sqrt{d}}\sum_{S_N} |j,j\rangle_L\otimes 
|{\bar j},{\bar j}\rangle_R
\eea
Since $SU(2)$ Young diagrams with $N$ boxes can also be viewed as $S_N$ Young diagrams, and different representations of $S_N$ are orthogonal, the sum in (\ref{OrbYoung}) gives zero unless 
\bea
j={\bar j}.
\eea
To illustrate this general property, we give an explicit example of $N=2$, $j=1$, ${\bar j}=0$:
\bea
|0,0;1,1\rangle=\frac{1}{\sqrt{2}}
\left\{\frac{1}{\sqrt{2}}\left[|\uparrow\downarrow\rangle-|\downarrow\uparrow\rangle\right]_L\otimes |\uparrow\uparrow\rangle_R+
\frac{1}{\sqrt{2}}\left[|\da\ua\rangle-|\ua\da\rangle\right]_L\otimes |\uparrow\uparrow\rangle_R
\right\}=0.\nonumber
\eea

When (\ref{OrbYoung}) is multiplied by a particular state (\ref{SigmaToSpin}), only one term in the sum (\ref{OrbYoung}) contributes, so the large $N$ scaling of the product (which gives the correlation function (\ref{Tcorrel})) is determined by $d$. Recalling that $d$ is a dimension of the $S_N$ orbit, which is the same as the dimension of representation corresponding to a given Young diagram ($\frac{N}{2}+j$ boxes in the first row and $\frac{N}{2}-j$ boxes in the second row), we can evaluate this number using the hook--length formula:
\bea
d=\frac{N!}{(\frac{N}{2}-j)!(\frac{N}{2}+j+1)\dots (2j+2)
(N-2[\frac{N}{2}-j])!}=
\frac{(2j+1)N!}{(\frac{N}{2}-j)!(\frac{N}{2}+j+1)!}.
\eea 
Recalling that $N\gg \frac{N}{2}-j\sim 1$ and using Stirling's formula, we find
\bea
d\sim N^{\frac{N}{2}-j}.
\eea
This determines the scaling of (\ref{Tcorrel}) for the highest weight state $T_{j,j}$:
\bea\label{YoungSeFin}
\left\langle S_{N,0}(v)T_{j;j}^\dagger(w)\left((J_0^-)^p
({\bar J}_0^-)^{\bar p}\sigma_n^P\right)(0)
\left((J_0^+)^q
({\bar J}_0^+)^{\bar q}\sigma_n^{Q\dagger}\right)(a)
\right\rangle_{S_N}
%\left\langle S_{N,0}(v)T_{j,j}^\dagger(w)\left((J_0^-)^p\sigma_n^P\right)(0)
%\left((J_0^+)^q\sigma_n^{Q\dagger}\right)(a)\right
%\rangle_{S_N}
\sim \frac{1}{\sqrt{d}}\sim 
N^{-(\frac{N}{2}-j)/2}
\eea
and proves the formula (\ref{FinConj}) for 
$k={\bar k}=\frac{N}{2}-j$. In the next subsection we will extend this result to other values of $k$ and ${\bar k}$.

\subsection{Other members of the multiplet}
\label{SecLrgN2}

We will now allow the final state $T$ to be an arbitrary member of the multiplet:
\bea\label{OrbYComb}
&&\left|j,\frac{N}{2}-k;{\bar j},\frac{N}{2}-{\bar k}\right\rangle=
\sqrt{\frac{(2j-l)!}{(2j)!l!}}
\sqrt{\frac{(2{\bar j}-{\bar l})!}{(2{\bar j})!{\bar l}!}}
(J_0^-)^l({\bar J}_0^-)^{\bar l}|j,j;{\bar j},{\bar j}\rangle\\
&&l\equiv j+k-\frac{N}{2},\qquad 
{\bar l}\equiv {\bar j}+{\bar k}-\frac{N}{2},\nonumber\\
&&|j,j;{\bar j},{\bar j}\rangle=\frac{1}{\sqrt{d}}\sum_{S_N} |j,j\rangle_L\otimes 
|{\bar j},{\bar j}\rangle_R\nonumber
\eea
As we demonstrated above, summation in the last line leads to the restriction $j={\bar j}$. The amplitude (\ref{Tcorrel}) is obtained by multiplying (\ref{OrbYComb}) and an appropriate linear combination of (\ref{SigmaToSpin}). Then by moving $(J_0^-)$, $({\bar J}_0^-)$ to (\ref{SigmaToSpin}), such amplitude is reduced to a combination of the results with $l={\bar l}=0$. Although this reduction involves some nontrivial combinatorics, 
for our purposed it is sufficient to notice that all combinatorial coefficients scale as $N^0$ since $J^-_0$, ${\bar J}_0^-$ act only on the first $n$ copies and $n\sim 1$. This implies that the $N$ scaling in (\ref{FinConj}) is determined by $d$, which was evaluated in the last subsection, and by factorials in (\ref{OrbYComb}). 

For small values of $l$, we approximate the factorials in (\ref{OrbYComb}),
\bea
\left|j,\frac{N}{2}-k;{j},\frac{N}{2}-{\bar k}\right\rangle
&\sim& N^{-(l+{\bar l})/2}
(J_0^-)^l({\bar J}_0^-)^{\bar l}|j,j;{j},{j}\rangle\nonumber\\
&\sim& N^{-(k+{\bar k})/2}N^{\frac{N}{2}-j}
(J_0^-)^l({\bar J}_0^-)^{\bar l}|j,j;{j},{j}\rangle
\eea
and combining this with (\ref{YoungSeFin}), we arrive at (\ref{FinConj}).

\end{document}